\documentclass{article}

\usepackage[preprint]{neurips_2025}
\setcitestyle{numbers,square}

\usepackage{microtype}
\usepackage{enumitem}
\usepackage{graphicx}
\usepackage{subcaption}
\usepackage{adjustbox} 
\usepackage{placeins}

\usepackage[utf8]{inputenc} 
\usepackage[T1]{fontenc}    
\usepackage{hyperref}       
\usepackage{url}            
\usepackage{booktabs}       
\usepackage{tabularx} 
\usepackage{multirow} 
\usepackage{array} 
\usepackage{amsfonts}       
\usepackage{nicefrac}       
\usepackage{threeparttable} 
\usepackage{xcolor}         

\usepackage{amsmath}
\usepackage{amssymb}
\usepackage{mathtools}
\usepackage{amsthm}

\usepackage[capitalize,noabbrev]{cleveref}

\theoremstyle{plain}

\theoremstyle{definition}

\theoremstyle{remark}

\newcommand{\bbP}{\mathbb{P}}
\newcommand{\bbE}{\mathbb{E}}

\newcommand{\E}{\bbE}                                   
\renewcommand{\Pr}{\bbP}                                

\title{Controllable Generative Sandbox for Causal Inference}

\author{%
  Qi Zhang \\
  Emory University\\
  \texttt{qi.zhang2@emory.edu} \\
  \And
  Harsh Parikh \\
  Yale University\\
  \texttt{harsh.parikh@yale.edu} \\
  \AND
  Ashley Naimi \\
  Emory University\\
  \texttt{ashley.naimi@emory.edu} \\
  \And
  Razieh Nabi \\
  Emory University \\
  \texttt{razieh.nabi@emory.edu} \\
  \And
  Christopher Kim \\
  Amgen\\
  \texttt{chrkim@amgen.com} \\
  \And
  Timothy Lash \\
  Emory University\\
  \texttt{timothy.lee.lash@emory.edu} \\
}

\begin{document}

\maketitle

\begin{abstract}
Method validation and study design in causal inference rely on synthetic data with known counterfactuals. Existing simulators trade off \emph{distributional realism}, the ability to capture mixed-type and multimodal tabular data, against \emph{causal controllability}, including explicit control over overlap, unmeasured confounding, and treatment effect heterogeneity. We introduce \textsc{CausalMix}, a variational generative framework that closes this gap by coupling a mixture of Gaussian latent priors with data-type-specific decoders for continuous, binary, and categorical variables. The model incorporates explicit causal controls: an overlap regularizer shaping propensity-score distributions, alongside direct parameterizations of confounding strength and effect heterogeneity. This unified objective preserves fidelity to the observed data while enabling factorial manipulation of causal mechanisms, allowing overlap, confounding strength, and treatment effect heterogeneity to be varied independently at design time. Across benchmarks, \textsc{CausalMix} achieves state-of-the-art distributional metrics on mixed-type tables while providing stable, fine-grained causal control. We demonstrate practical utility in a comparative safety study of metastatic castration-resistant prostate cancer treatments, using \textsc{CausalMix} to compare estimators under calibrated data-generating processes, tune hyperparameters, and conduct simulation-based power analyses under targeted treatment effect heterogeneity scenarios.

\end{abstract}

\section{Introduction}

For each unit, only one potential outcome is observed, rendering individual causal effects fundamentally unobservable \citep{holland1986statistics}. Consequently, assessing causal methods requires settings in which the ground truth is known.
\emph{Synthetic data} therefore serves as a workhorse for causal methodology: it supplies ground-truth counterfactuals, standardized benchmarks, and controllable stress-tests of identification assumptions. In this \textit{sandbox}, researchers can probe robustness to confounding, model misspecification, and limited overlap; characterize finite-sample behavior; and tune hyperparameters (e.g., causal forest depth, super-learner libraries and cross-fitting folds) under conditions that mirror the intended application—all while retaining access to the truth \citep{hahn2019atlantic,mahajan2022empirical}. Real data-inspired simulations also support prospective study design (power and sample size under targeted heterogeneity/confounding) and privacy-preserving collaboration by releasing realistic, non-identifiable surrogates \citep{pezoulas2024synthetic}.

Performance on synthetic data often fails to translate to complex real-world settings because estimator behavior depends on how well modeling assumptions align with properties of the data, including dependence structure, nonlinearity, confounding strength, overlap, outcome variability, and effect size. Consequently, an ideal benchmark must \emph{both} provide access to (causal) truth \emph{and} faithfully reproduce key features of the real data-generating process (DGP) \citep{friedrich2024role,gamella2025physical}. 

\textbf{Gaps.}
Despite significant advances in deep generative architectures for causal data \citep{neal2020realcause,athey2024using,parikh2022validating}, existing approaches continue to face a fundamental challenge: jointly achieving realistic modeling of mixed-type observational data and explicit, design-time control over key causal properties such as effect heterogeneity, confounding, and overlap. As a result, gains in empirical realism often come at the expense of causal transparency, limiting their usefulness for method evaluation and study planning.

\textbf{Contributions.} We propose \textsc{CausalMix}, a variational generative framework that unifies distributional realism with fine-grained causal control.
\begin{itemize}[leftmargin=*]
\item \textbf{Mixed-type fidelity.} A hybrid latent architecture with a Bayesian Gaussian--mixture prior and data-type-specific decoder heads captures multimodal dependencies common in observational and clinical datasets.
\item \textbf{Causal levers.} We expose \emph{design-time} controls for heterogeneous treatment effects, propensity overlap, and unmeasured confounding. Overlap is tuned via a log-likelihood--ratio regularizer that directly shapes treatment assignment; confounding and effect heterogeneity are parameterized in the outcome mechanism.
\item \textbf{Stabilized causal fidelity.} Regularization and variance constraints ensure that prespecified causal functions and controls are faithfully realized during training, particularly when causal mechanisms are low-dimensional or weakly nonlinear.
\item \textbf{Unified objective.} The model jointly optimizes distributional fit and causal constraints, ensuring that matching the observed data does not compromise control over the specified causal structure. 
\end{itemize}

We further introduce an integrated evaluation and application pipeline for causal synthetic data that jointly assesses distributional fidelity, causal fidelity, and privacy in terms of record-level disclosure risk under synthetic data release. Practitioners need calibrated sandboxes that both \emph{look like their data} and permit targeted “what-if’’ studies about overlap, confounding, and heterogeneity. \textsc{CausalMix} enables (i) apples-to-apples \emph{estimator comparison} under empirically grounded DGPs, (ii) principled \emph{hyperparameter selection} within the same calibrated regime, and (iii) \emph{simulation-based study design} (power/sample size calculation under tunable heterogeneity and residual confounding). We illustrate these benefits in a comparative-safety setting (abiraterone vs.\ enzalutamide in metastatic castration-resistant prostate cancer), where observational data exhibit realistic structure and plausible biases. The code is publicly available at \url{https://github.com/zhangqiecho/causalmix}.
\section{Background}
\subsection{Setup and Notation}
We observe i.i.d.\ samples $O=(X,T,Y)$, where $X\in\mathcal{X}$ denotes covariates, $T\in\{0,1\}$ a binary treatment, and $Y$ the observed outcome. Let $Y(1)$ and $Y(0)$ be the potential outcomes,
with $Y=T\,Y(1)+(1-T)\,Y(0)$ by consistency. We define the conditional outcome regression $\mu_t(x) \triangleq \E[Y(t)\mid X=x]$ and the propensity score $e(x) \triangleq \Pr(T=1\mid X=x)$ \citep{rosenbaum1983central}, and consider the average and conditional average treatment effects $\text{ATE}=\E[Y(1)-Y(0)]$ and $\text{CATE}(x)=\mu_1(x)-\mu_0(x)$.

Identification of average and conditional treatment effects from observational data typically relies on assumptions of consistency, unconfoundedness given observed covariates $X$, and positivity. In practice, however, the finite-sample behavior of causal estimators depends strongly on the strength of confounding, model complexity, and the degree of propensity-score overlap. Violations of ignorability due to unmeasured confounders, as well as limited overlap between treatment groups, undermine point identification, inflate variance, and forces extrapolation of counterfactual outcomes in regions with limited overlap, complicating both estimation and method comparison.

In such settings, synthetic data with a known and controllable causal structure provides a principled testbed to study these challenges: it enables systematic control over unmeasured confounding and overlap, supports diagnostic evaluation through known ground truth, and facilitates fair benchmarking of causal estimators under realistic departures from ideal identification conditions.

\subsection{Background on Synthetic Data for Causal Inference}
Synthetic data underpins algorithm development, fair comparison, and study design when counterfactual truths are otherwise unobservable \citep{holland1986statistics}. We summarize three families, ordered by how tightly they are grounded in real data.

\textbf{Arbitrary parametric simulators.}
Classical setups specify structural equations for $(X,T,Y)$, granting full control over effect sizes, functional forms, and noise. They are easy to vary and analyze but can yield method rankings that reflect the designer’s choices more than real-world complexity, especially for mixed-type, multimodal tabular data.

\textbf{Semi-synthetic benchmarks.}
These approaches reuse covariates $X$ from real datasets while simulating $T\!\mid\!X$ and $Y\!\mid\!T,X$ from fixed mechanisms. This paradigm underlies widely used challenges and leaderboards \citep{shimoni2018benchmarking,hahn2019atlantic,dorie2019automated,NEURIPS2021_8526e096}, enabling reproducible comparisons but offering limited, indirect control over causal “knobs’’ such as overlap and heterogeneity, with potential privacy constraints in sensitive domains.

\textbf{Data-fit generators.}
A more recent direction learns the synthetic DGP from real data so that marginal and conditional distributions match empirical structure.
\emph{Plasmode} simulations preserve observed $X$ and (sometimes) $T$, while generating outcomes from known models to inject ground truth \citep{abadie2011bias,franklin2014plasmode,liu2019missing,ress2024comparing}; they are transparent but can be inflexible and sensitive to design choices \citep{cattell1967general,gadbury2008evaluating,vaughan2009use,shaw2025cautionary}.
Neural approaches fit $T\!\mid\!X$ (or $X\!\mid\!T$) and $Y\!\mid\!T,X$ using flexible generative models, including normalizing flows in \textsc{RealCause} \citep{neal2020realcause}, Wasserstein Generative Adversarial Networks (WGANs) \citep{athey2024using}, and conditional variational autoencoder (VAE) frameworks \citep{parikh2022validating,boulet2025draw}. These models improve distributional fidelity but may suffer from training instability (for GANs) \citep{arjovsky2017wasserstein}, limited support for mixed-type tabular data \citep{parikh2022validating}, or only coarse and indirect control over overlap, unmeasured confounding, and effect heterogeneity \citep{neal2020realcause}.

Methods emphasizing empirical realism, such as \textsc{RealCause} and WGAN-based generators, closely mimic observed data distributions but offer limited explicit control over causal parameters. For instance, \textsc{RealCause} adjusts effect scale, heterogeneity, and overlap via interpolation between fitted extremes while assuming no unmeasured confounding \citep{neal2020realcause}, and WGAN-based approaches lack direct mechanisms for manipulating treatment effects \citep{athey2024using}. 
Frengression \citep{yang2025frugal} enables direct control over marginal causal quantities through interventional modeling, but does not allow explicit specification of conditional effects, overlap, or unmeasured confounding, which are central to evaluating estimator robustness in observational settings. \textsc{Credence} \citep{parikh2022validating} allows users to prespecify functional forms for treatment effects and unmeasured confounding bias, but is limited in modeling realistic mixed-type, multimodal tabular data. Moreover, it lacks systematic empirical validation that the learned generator actually satisfies the prespecified causal functions and confounding controls.

\textbf{Desiderata for benchmarks.}
Across applications, three properties are especially valuable: 
(i) \textbf{mixed-type realism}—faithfully modeling continuous, binary, and categorical variables with multimodality; 
(ii) \textbf{causal controllability}—treating effect size and heterogeneity, unmeasured confounding, and overlap as explicit \emph{design parameters}; and 
(iii) \textbf{joint optimization}—achieving distributional fidelity and causal control without trading one for the other. 
These desiderata motivate the framework developed in this paper, which extends conditional VAE–based generators with mixture priors, datatype-specific decoders, and explicit treatment effect, overlap and confounding controls.

\section{Methods}
\subsection{Problem Formulation}
The primary goal of this work is to learn a flexible data-generating mechanism
$G_\theta$ such that the joint distribution of a generated sample, $\mathcal{O}'=(X',T',Y') \sim G_\theta$ closely approximates the empirical distributions of the observed data, $p_{\text{real}}(X,T,Y)$, while also providing explicit and user-controlled manipulation of key causal
properties. To support this goal, we adopt a nonparametric structural causal model (SCM)
in which all structural equations are unrestricted mappings.

Let the observed sample
$\mathcal{O}=\{(X_i,T_i,Y_i)\}_{i=1}^n$
be generated from a latent causal process
$G^{\dagger} = [\phi_X,\phi_T,\phi_{Y(1)},\phi_{Y(0)}]$
specified by:

\vspace{-0.5cm}
\begin{align*}
X &= \phi_X(U_X),\\
T &= \phi_T(X, U_T),\\
Y(t) &= \phi_{Y(t)}(X, U_Y), \quad t\in\{0,1\}, \\
Y &= T\,Y(1) + (1-T)\,Y(0), 
\end{align*}
where $U_X$, $U_T$, and $U_Y$ are mutually independent exogenous variables, and each $\phi_{\cdot}$ is an arbitrary, nonlinear function.  
This SCM induces the true joint distribution $p_{\text{real}}(X,T,Y)$ that our learned generator aims to approximate.

\textbf{Causal control functions.}
Beyond learning distributions that match observed data, we require the ability to specify key causal parameters in the generated synthetic data. We formalize three dimensions of causal control by introducing a set of user-specified, covariate-dependent control functions:
\[
\Psi = \{\alpha(X),\, \tau(X),\, \kappa(X,T)\}.
\]

Each function governs a distinct causal property within the SCM:
\begin{itemize}[leftmargin=*]
    \item \textbf{Overlap} refers to how similar the covariate distributions are for treated $P(X\mid T=1)$ vs. untreated groups $P(X\mid T=0)$. To specify the desired positivity/overlap structure, we introduce a covariate-dependent
    \emph{conditional density ratio}
    \[
    \alpha(x)=
    \frac{P(X = x \mid T=0)}{P(X = x\mid T=1)}.
    \]
    Values of $\alpha(X)$ near one indicate strong overlap between treatment groups, whereas large
    or small values correspond to regions of weak overlap or near-positivity violations in which the
    covariate space is dominated by the untreated or treated group, respectively.

    \item \textbf{Effect function} $\tau(X)$ defines the CATE:
    \[
    \tau(x) = \mathbb{E}[Y(1)-Y(0)\mid X=x], 
    \]
    determining both the \emph{magnitude} and \emph{heterogeneity} of causal effects across covariates.
    \item \textbf{Unmeasured confounding} $\kappa(X,T)$ introduces a dependence between $T$ and the potential outcomes $Y(t)$ through shared latent factors, capturing the degree and form of hidden confounding beyond $X$. It is defined as:
    \begin{equation*}
    \begin{split}
    \kappa(x,t) 
    &= \mathbb{E}[Y(t)\mid X=x, T=1]  - \mathbb{E}[Y(t)\mid X=x, T=0].
    \end{split}
    \end{equation*}
\end{itemize}
$\kappa(x,t)$ encodes structural unmeasured confounding through its induced selection bias in potential outcomes across treatment groups, thereby governing departures from conditional ignorability.

Recent developments in semiparametric causal inference have expanded identification and estimation theory to complex directed acyclic graphs with latent variables, enabling consistent estimation of causal functionals beyond classical back-door and front-door settings \citep{guo2023flexible, guo2024average, guo2025causal}. These approaches employ flexible machine-learning estimators together with one-step corrections or targeted minimum loss procedures to obtain doubly robust and asymptotically efficient estimates under weak modeling assumptions. However, their practical reliability depends critically on properties of the underlying data-generating process, including overlap, nuisance complexity, and the interaction between treatment assignment and latent confounding. Because such factors cannot be systematically varied in observational datasets, principled evaluation requires a generator capable of independently controlling these causal mechanisms. The control functions introduced above are designed precisely to support this type of estimator-level stress testing.

\textbf{Joint optimization problem.}
Given the causal control functions $\Psi=\{\alpha,\tau,\kappa\}$, we want to learn a data generator
$G_\theta$ whose induced distribution over generated samples
$\mathcal{O}'=(X',T',Y') \sim G_\theta$
is as close as possible to the empirical joint distribution $p_{\text{real}}(X,T,Y)$ while
remaining faithful—up to user-controlled rigidness—to the causal structure encoded in $\Psi$.

For any generator $G_\theta$, we define the induced causal quantities using expectations over
generated samples:
\[
\begin{aligned}
\alpha_\theta(x)
&:= \frac{p_\theta(X=x \mid T'=0)}{p_\theta(X=x \mid T'=1)}, \\[6pt]
\tau_\theta(x)
&:= \mathbb{E}[Y'(1) - Y'(0) \mid X=x], \\[6pt]
\kappa_\theta(x,t)
&:= \mathbb{E}[Y'(t) \mid X=x,T=1] - \mathbb{E}[Y'(t) \mid X=x,T=0].
\end{aligned}
\]

The subset of generators that satisfy the causal specifications exactly is

\[
\mathcal{G}^*_\Theta(\Psi)
:= \left\{\, G_\theta :\;
\begin{aligned}
\alpha_\theta(X) &= \alpha(X), \\
\tau_\theta(X) &= \tau(X), \\
\kappa_\theta(X,T) &= \kappa(X,T)
\end{aligned}
\right\}.
\]

Similarly, let
\[
\mathcal{G}^{\dagger}_\Theta
:= \{G_\theta : p_\theta(X',T',Y') \approx p_{\text{real}}(X,T,Y)\}
\]
be the set of generators reproducing the empirical distribution.  
If $\mathcal{G}^{\dagger}_\Theta \cap \mathcal{G}^{*}_\Theta(\Psi)\neq\emptyset$, an ideal solution
lies in this intersection; otherwise, we seek a generator in $\mathcal{G}^*_\Theta(\Psi)$ that is
closest to $\mathcal{G}^{\dagger}_\Theta$ under a chosen discrepancy.

This leads to the optimization principle:

\[
\begin{aligned}
\min_{\theta}\;
&\mathbb{E}\!\left[d\!\big((X,T,Y),(X',T',Y')\big)\right] \\
&\quad + \lambda_\alpha\,\mathbb{E}_{X}\!\left[
    \mathcal{R}_\alpha(\alpha_\theta(X),\alpha(X))
\right] \\
&\quad + \lambda_\tau\,\mathbb{E}_{X}\!\left[
    \mathcal{R}_\tau(\tau_\theta(X),\tau(X))
\right] \\
&\quad + \lambda_\kappa\,\mathbb{E}_{X,T}\!\left[
    \mathcal{R}_\kappa(\kappa_\theta(X,T),\kappa(X,T))
\right].
\end{aligned}
\]

where $d(\cdot,\cdot)$ is a distributional distance, 
$\mathcal{R}_{\alpha},\mathcal{R}_{\tau},\mathcal{R}_{\kappa}$ are rigidness penalties that regulate adherence to the overlap, treatment-effect, and confounding specifications, and $\lambda_\alpha, \lambda_\tau, \lambda_\kappa$ are regularization parameters. 

The first term enforces fidelity to the empirical distribution, whereas the penalty terms guide
$G_\theta$ toward the user-specified causal structure.  This framework learns a data-generating
mechanism that approximates the true observational process while enabling controlled variation
along three interpretable causal dimensions.

\textbf{Problem statement}: Our goal is to develop a model architecture, training algorithm, and data generation procedure that:
\begin{itemize}[leftmargin=*]
    \item Effectively optimizes the unified training objective to find parameter values that balance all objectives
    \item Satisfies the causal constraints with high fidelity (constraints are met within acceptable tolerance)
    \item Handles mixed-type tabular data without preprocessing artifacts
    \item Provides interpretable control over causal parameters through easily tunable hyperparameters
    \item Generates high-quality synthetic samples efficiently
\end{itemize}

The following sections present our proposed solution to this problem formulation, introducing specific design choices for the model architecture, prior distribution, causal regularization mechanism, and training procedures.

\subsection{Model Overview and Data-Generating Process}
\begin{figure*}[t]
\centering
\includegraphics[width=\textwidth]{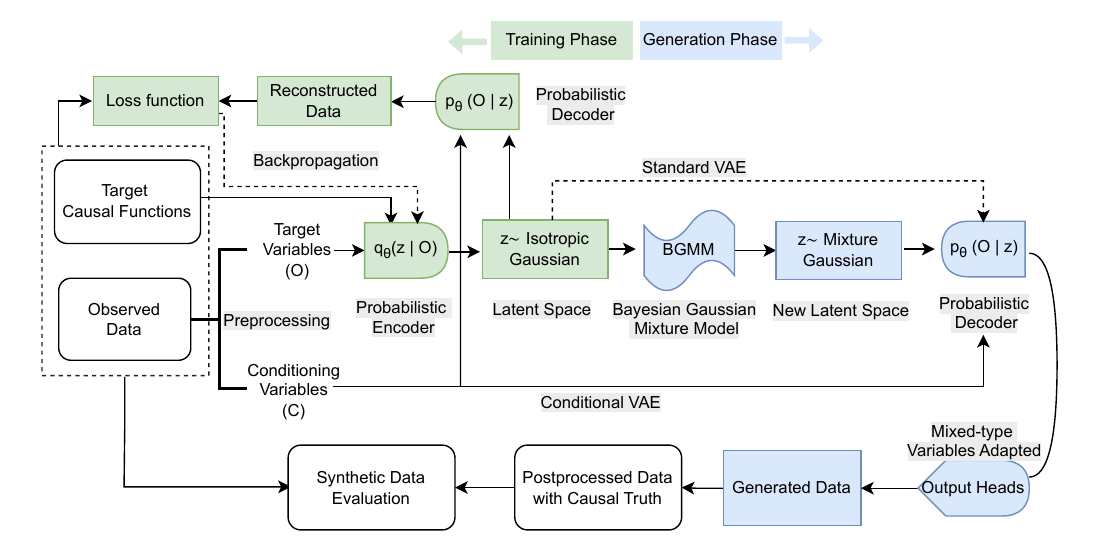}
\caption{Overview of the \textsc{CausalMix} generative framework.}
\label{fig:diagram}
\end{figure*}

\textsc{CausalMix} instantiates a modular data-generating mechanism $G_\theta$ built on a conditional VAE backbone. The observational distribution is factorized as
\[
p(X,T,Y) = p(T)\,p(X\mid T)\,p(Y\mid X,T),
\]
and modeled using three components: a treatment model for $T$, a pre-treatment generator for $X\mid T$, and a post-treatment generator for the potential outcomes $(Y(0),Y(1))\mid X,T$.

\paragraph{Modular generators.}
Given variable metadata, the \textsc{CausalMix} wrapper handles preprocessing and instantiates:
(i) a Bernoulli model for treatment assignment,
(ii) a conditional VAE $G_{X,\theta}$ for $X\mid T$, and
(iii) a conditional VAE $G_{Y,\theta}$ for $(Y(0),Y(1))\mid X,T$.
Synthetic data are generated sequentially by sampling $T'$, then $X'\mid T'$, and finally $(Y'(0),Y'(1))\mid X',T'$, with the observed outcome constructed as
$Y' = T'Y'(1) + (1-T')Y'(0)$. Generated samples are deterministically transformed back to the original data scale and representation using the preprocessing metadata (e.g., inverse standardization, re-integerization and categorical decoding).

\paragraph{Conditional VAE backbone.}
Both generators share a common conditional VAE structure with multilayer perceptron (MLP) encoders and decoders. The encoder maps target variables to a latent representation and outputs the mean and log-variance of a diagonal Gaussian posterior. The decoder maps the concatenated latent code and conditioning variables (either $T$ or $(X,T)$) to a shared hidden representation, which is then routed through variable-specific output heads for binary, categorical, and continuous variables, enabling realistic modeling of mixed-type tabular data.

\paragraph{Potential outcomes and causal constraints.}
The post-treatment generator $G_{Y,\theta}$ jointly models both potential outcomes, enabling direct access to individual-level treatment effects. During training, reconstruction is conditioned on the observed treatment arm, while additional regularization enforces alignment with user-specified treatment-effect and confounding functions. Stabilization mechanisms are introduced to ensure that these causal targets are faithfully realized; we describe the corresponding objective in \cref{sec:objective}.

\paragraph{Overlap control.}
The covariate generator $G_{X,\theta}$ incorporates an explicit overlap control mechanism that shapes the conditional distribution $p(X\mid T)$ to match a user-specified log-density ratio. This provides direct, design-time control over propensity-score overlap; implementation details are deferred to \cref{sec:objective}.

\paragraph{Latent prior.}
A key design choice in \textsc{CausalMix} is the use of a flexible \emph{mixture-based latent prior} to capture multimodal structure in observational data. Instead of restricting the latent space to a standard Gaussian, each conditional VAE can employ a Bayesian Gaussian mixture model (BGMM) fitted in latent space after training. Latent samples drawn from this mixture prior are then decoded conditionally to generate $X'$ and $(Y'(0),Y'(1))$, improving the model’s ability to represent complex, heterogeneous distributions.

Overall, this modular design enables \textsc{CausalMix} to generate synthetic datasets that preserve empirical structure while exposing explicit controls over overlap, unmeasured confounding, and treatment-effect heterogeneity. Figure~\ref{fig:diagram} provides an overview of the \textsc{CausalMix} framework. 

\subsection{Objective Function}
\label{sec:objective}

The parameters $\theta$ of the generative mechanism $G_\theta$ are learned by optimizing a unified objective that combines generative modeling with penalties enforcing user-specified causal structure. The objective integrates three components: (i) a variational autoencoder loss for distributional fidelity, (ii) penalties aligning induced causal quantities with target causal functions, and (iii) an overlap penalty controlling the covariate distribution $p_\theta(X\mid T)$. We describe each component in turn.

\textbf{Generative modeling objective.}
Each module in \textsc{CausalMix} is implemented as a conditional VAE. For a batch of observed data $(X,T,Y)$ and latent variables $Z \sim q_\theta(z\mid\cdot)$, reconstruction is performed using likelihoods appropriate to each variable type: Gaussian likelihoods for continuous variables (Gaussian negative log-likelihood, rather than the MSE used in \textsc{Credence}), Bernoulli likelihoods for binary variables (logistic cross-entropy), and categorical likelihoods for discrete variables (softmax cross-entropy).

Training maximizes the evidence lower bound (ELBO). For observed variables $\mathcal{O}$ (either $X$ or $Y$), the ELBO is
\[
\mathrm{ELBO}(\theta)
=
\mathbb{E}_{q_\theta(z)}
\big[
\log p_\theta(\mathcal{O}\mid z)
\big]
-
\mathrm{KL}\big(q_\theta(z)\,\|\,p(z)\big),
\]

where $p(z)$ denotes the latent prior. Equivalently, minimizing the negative ELBO yields

\[
-\mathrm{ELBO}(\theta)
=
-
\mathbb{E}_{q_\theta(z)}
\big[
\log p_\theta(\mathcal{O}\mid z)
\big]
\;+\;
\mathrm{KL}\big(q_\theta(z)\,\|\,p(z)\big).
\]

The encoder parameterizes a diagonal Gaussian posterior
$q_\theta(z)=\mathcal{N}(\mu_\theta,\mathrm{diag}(\sigma_\theta^2))$, for which the KL divergence admits the closed form
\[
\mathrm{KL}\big(q_\theta(z)\,\Vert\,p(z)\big)
=
\frac{1}{2}
\sum_{j=1}^{d}
\big(
\mu_{\theta,j}^2
+
\sigma_{\theta,j}^2
-
\log \sigma_{\theta,j}^2
-
1
\big).
\]

Combining the reconstruction and KL terms gives the standard VAE loss
\[
\mathcal{L}_{\mathrm{VAE}}(\theta)
=
-\mathbb{E}_{q_\theta(z)}
\big[
\log p_\theta(\mathcal{O}\mid z)
\big]
\;+\;
\lambda_{\mathrm{KL}}\,
\mathrm{KL}\big(q_\theta(z)\,\|\,p(z)\big),
\]
where $\lambda_{\mathrm{KL}}$ controls the strength of latent regularization. In the post-treatment generator, reconstruction is conditioned on the observed treatment arm and uses only the corresponding outcome $Y(T)$.

\textbf{Causal parameter control.}
To ensure that the learned generator realizes the desired causal structure, we introduce penalties on induced causal quantities defined by user-specified control functions:
$\alpha(X)$ for overlap,
$\tau(X)$ for treatment-effect magnitude and heterogeneity, and
$\kappa(X,T)$ for unmeasured confounding.
The strength of each constraint is governed by rigidness parameters
$\lambda_\alpha,\lambda_\tau,\lambda_\kappa\ge 0$.

\textbf{Treatment-effect control.}
To align the learned treatment-effect function with its target, $\tau_\theta(X)\approx\tau(X)$, we use a weighted composite penalty \text{(Huber loss)} that shapes the distribution of CATE residuals $\Delta\tau_\theta(X)=\tau_\theta(X)-\tau(X)$. The quadratic term anchors mean alignment of these residuals, implicitly stabilizing the induced ATE, while the Smooth~L1 term improves robustness to outliers and weakly identified regions, trading off precise alignment against stability when the target function is low-dimensional or weakly nonlinear. A separate variance regularizer discourages spurious unit-level dispersion around the target causal structure, stabilizing training without collapsing intended heterogeneity.

\begin{equation*}
\mathcal{L}_{\tau}^{\mathrm{mean}}(\theta)
=
\mathbb{E}_{X}\!\Big[
\lambda_{\tau}^{\mathrm{mse}} (\Delta\tau_\theta)^2
+
\lambda_{\tau}^{\mathrm{sl1}}\,
\mathrm{SmoothL1}(\Delta\tau_\theta)
\Big],
\quad \text{and} \quad
\mathcal{L}_{\tau}^{\mathrm{var}}(\theta)
=
\mathrm{Var}\!\big[\Delta\tau_\theta\big].
\end{equation*}

\textbf{Unmeasured confounding control.}
Let $\kappa_\theta(X,T)$ denote the confounding bias induced by the outcome decoder. We impose an analogous composite penalty on the confounding residuals $\Delta\kappa_\theta(X,T)=\kappa_\theta(X,T)-\kappa(X,T)$, using quadratic and Smooth~L1 terms for mean alignment and robustness, together with a variance regularizer to control excess unit-level dispersion:

\begin{equation*}
\mathcal{L}_{\kappa}^{\mathrm{mean}}(\theta)
=
\mathbb{E}_{X,T}\!\Big[
\lambda_{\kappa}^{\mathrm{mse}} (\Delta\kappa_\theta)^2
+
\lambda_{\kappa}^{\mathrm{sl1}}\,
\mathrm{SmoothL1}(\Delta\kappa_\theta)
\Big],
\quad \text{and} \quad
\mathcal{L}_{\kappa}^{\mathrm{var}}(\theta)
=
\mathrm{Var}\!\big[\Delta\kappa_\theta\big].
\end{equation*}

\textbf{Overlap control.}
A key feature of \textsc{CausalMix} is direct control over covariate overlap between treatment groups through the log-density ratio

\[
\log \alpha_\theta(X)
=
\log p_\theta(X \mid T'=0)
-
\log p_\theta(X \mid T'=1),
\]

which quantifies the relative likelihood of a covariate profile under treatment versus control. By penalizing deviations of $\log \alpha_\theta(X)$ from a user-specified target $\log\alpha(X)$, the generator can induce regimes ranging from strong overlap to near-positivity violations.

The conditional log-densities $\log p_\theta(X\mid T'=t)$ are computed from feature-wise decoder likelihoods. Binary, categorical, and continuous variables contribute Bernoulli, softmax, and Gaussian log-densities, respectively, as detailed in \cref{app:llr}. The overlap penalty takes the form
\[
\mathcal{L}_{\alpha}(\theta)
=
\mathbb{E}_{X}\!\left[
\big(
\log \alpha_\theta(X)
-
\log \alpha(X)
\big)^2
\right],
\]
with $\lambda_\alpha$ controlling the rigidity of the constraint.

\textbf{Unified objective.}
The full learning objective combines distributional fidelity with explicit causal control:
\[
\begin{aligned}
\mathcal{L}(\theta)
&=
\mathcal{L}_{\mathrm{VAE}}(\theta)
+\lambda_\alpha\, \mathcal{L}_{\alpha}(\theta)
+\lambda_\tau\, \mathcal{L}_{\tau}^{\mathrm{mean}}(\theta)
+\lambda_{\tau}^{\mathrm{var}}\, \mathcal{L}_{\tau}^{\mathrm{var}}(\theta) \\
&\quad
+\lambda_\kappa\, \mathcal{L}_{\kappa}^{\mathrm{mean}}(\theta)
+\lambda_{\kappa}^{\mathrm{var}}\, \mathcal{L}_{\kappa}^{\mathrm{var}}(\theta).
\end{aligned}
\]

Minimizing $\mathcal{L}(\theta)$ yields a generator that approximates the observational distribution while faithfully realizing the user-specified overlap, treatment-effect, and confounding structure.

\subsection{Mixed Variable Types}

Real-world tabular data typically contain heterogeneous variable types, including continuous,
binary, categorical, and integer-valued features. To accommodate this heterogeneity within a
unified generative framework, each conditional VAE module employs a multi-head decoder, with
each variable assigned a likelihood family consistent with its data type.

Binary variables are modeled using Bernoulli likelihoods parameterized by decoder logits.
Categorical variables are modeled using softmax logits with dimensionality matched to the
number of classes. Continuous variables are modeled using Gaussian likelihoods with
decoder-generated means and bounded log-variances, allowing the model to capture both
heteroscedasticity and variable-specific uncertainty. In our experiments, using Gaussian
negative log-likelihood for continuous variables, rather than mean-squared error as in
\textsc{Credence}, substantially improved performance, particularly for variables with
heterogeneous variance or constrained support. This improvement arises because the NLL
objective allows the decoder to learn both location and dispersion while maintaining
well-scaled gradients.

Integer-valued variables (e.g., counts stored as floats) are treated as continuous after
standardization, with optional rounding applied during post-processing. In potential-outcome
mode, each variable head emits separate parameter sets for $Y(0)$ and $Y(1)$; for numerical
stability, multiclass categorical outcomes are disallowed in this mode.

The exact per-variable log-likelihood expressions (Bernoulli, softmax, and Gaussian) used
during training and overlap computation are provided in \cref{app:llr}, where the total
log-density is formed by summing independent feature-wise contributions. Overall, this
multi-head decoder design enables accurate and flexible modeling of mixed-type observational
data within the VAE framework.

\subsection{Gaussian Mixture Model Prior}

Standard VAEs typically assume an isotropic Gaussian latent prior,
$p(z)=\mathcal{N}(0,I)$. However, recent work on VAE--GMM integration for tabular data
\citep{apellaniz2024improved} shows that real-world datasets often induce latent
representations that deviate substantially from a unimodal Gaussian due to complex,
nonlinear feature dependencies. Sampling from a standard Gaussian prior can therefore
distort the geometry of the learned latent space and reduce the realism of generated
samples, a limitation that is especially pronounced in heterogeneous tabular domains
characterized by multimodality and clustering.

To address this, \textsc{CausalMix} replaces the fixed Gaussian prior with a
\emph{Bayesian Gaussian mixture model} (BGMM) fitted post hoc to the encoder’s latent
means after VAE training. For each conditional VAE module, we collect encoder outputs
$\{\mu_\theta(\mathcal{O}_i)\}_{i=1}^N$ and fit a BGMM with a Dirichlet-process prior,
which adaptively infers the number of mixture components needed to represent the latent
distribution. The resulting prior takes the form
\[
p_{\mathrm{BGMM}}(z)
=
\sum_{k=1}^{K}
\pi_k\,\mathcal{N}(z \mid \mu_k,\Sigma_k),
\]
where $\pi_k$ denotes the mixture weight of component $k$, $\mu_k$ and $\Sigma_k$ are its mean and covariance, and $\sum_k \pi_k = 1$.
The Dirichlet-process prior encourages sparsity in the mixture weights, so that only a subset of components receives substantial probability mass. In practice, $K$ is learned
from data using truncated stick-breaking variational inference, with an upper bound set by the latent dimensionality. As emphasized in \citep{apellaniz2024improved}, this post hoc modeling preserves the original VAE training objective while providing a more faithful
approximation of the learned latent geometry.

During generation, latent samples are drawn from $p_{\mathrm{BGMM}}(z)$, optionally augmented with noise consistent with the average encoder variance, concatenated with conditioning features (e.g., $T'$ or $(X',T')$), and passed through the decoder trunk to produce a shared hidden representation from which variable-specific output heads generate model parameters.

The BGMM prior offers two key advantages. First, by modeling the latent space as a mixture
rather than a single Gaussian, it captures multimodal structure inherent in heterogeneous
tabular data, yielding more realistic synthetic samples. Second, it integrates cleanly with our causal control framework: while overlap, treatment effects, and unmeasured confounding are enforced at the decoder level, the mixture prior preserves latent flexibility and diversity without interfering with these constraints.

Together, these properties make the BGMM-enhanced VAE a central component of \textsc{CausalMix}, enabling high-quality synthetic data generation under both empirical and user-specified causal structure.

\subsection{Optimization Procedure}

Training proceeds by minimizing the unified objective described in \cref{sec:objective}, which combines the negative ELBO with penalties enforcing user-specified causal structure. Each conditional VAE
module is optimized independently using stochastic gradient descent with the Adam optimizer
(learning rate $1\times10^{-3}$). Models are trained using mini-batches drawn from an 80/20
train--validation split. We use PyTorch Lightning to manage batching, device placement, logging,
and early stopping.

\textbf{Forward pass and latent sampling.}
For each mini-batch, the encoder produces latent parameters $(\mu,\log\sigma^2)$, and latent
samples are drawn using the standard reparameterization trick $z = \mu + \sigma \odot \epsilon, \epsilon\sim\mathcal{N}(0,I).$
Conditioning features (either $T$ for $G_{X,\theta}$ or $(X,T)$ for $G_{Y,\theta}$) are concatenated with the latent code before decoding. Feature-specific output heads then produce logits for binary and categorical variables, and Gaussian parameters for continuous variables.

\textbf{Reconstruction and KL terms.}
The reconstruction loss is computed by summing per-variable negative log-likelihoods under the appropriate likelihood family, while the KL divergence uses the closed-form Gaussian--Gaussian
expression. For the post-treatment generator, reconstruction is conditioned on the observed treatment arm and includes only the corresponding potential outcome. 

\textbf{Causal penalties.}
During training of the post-treatment generator, the decoder also evaluates both potential
outcomes $Y'(0)$ and $Y'(1)$, allowing the induced treatment-effect function $\tau_\theta(X)$ and confounding function $\kappa_\theta(X,T)$ to be computed. Mean-alignment and variance penalties are applied as described in \cref{sec:objective} to enforce the corresponding causal constraints, with rigidity parameters controlling their relative strength. Similarly, the pre-treatment generator computes the per-sample log-density ratio
$\log p_\theta(X\mid T'=1)-\log p_\theta(X\mid T'=0)$, which is penalized against the target
overlap function $\log\alpha(X)$.

\textbf{Early stopping and monitoring.}
Validation loss is monitored during training, and early stopping is triggered when no improvement
is observed for several epochs. PyTorch Lightning logs the VAE loss, KL divergence, and individual
causal penalty terms separately, enabling diagnostics of both distributional fit and causal
constraint satisfaction.

\textbf{Post-training BGMM fitting.}
After convergence, the latent means $\mu_\theta(\mathcal{O}_i)$ from the training set are used to fit a Bayesian Gaussian mixture model with a Dirichlet-process prior. The maximum number of components is set equal to the latent dimensionality. This BGMM defines a flexible latent prior used during generation and does not participate in gradient-based optimization.

Together, these steps yield generators that approximate the empirical data distribution while
faithfully realizing the specified overlap, treatment-effect, and confounding structure.

\subsection{Evaluation Metrics}

We evaluate the proposed generators along three complementary axes:
(i) distributional fidelity between real and synthetic data,
(ii) causal-structure fidelity—including treatment-effect, confounding, and overlap behavior,
and (iii) privacy protection. All metrics compare the generated synthetic dataset to the observed data.
We use a combination of standard library-based and custom evaluation metrics; a summary of metric definitions and implementations is provided in \cref{tab:metric_summary}.

\textbf{Marginal fidelity.}
We assess one-dimensional marginals separately for continuous and discrete variables. For each
continuous feature $c$, we compute a normalized Wasserstein distance
\[
W_{\mathrm{norm}}(c)
=
\frac{W_1(F_{\mathrm{real}}^{(c)},F_{\mathrm{synth}}^{(c)})}
{\mathrm{sd}_{\mathrm{real}}^{(c)} + \varepsilon},
\]
where $F_{\mathrm{real}}^{(c)}$ and $F_{\mathrm{synth}}^{(c)}$ denote the empirical distributions of feature $c$ in the real and synthetic data, respectively, $W_1$ is the 1-Wasserstein distance, $\mathrm{sd}_{\mathrm{real}}^{(c)}$ is the standard deviation of feature $c$ in the
real data, and $\varepsilon>0$ is a small constant for numerical stability. We also report KSComplement, defined as
$1 - D_{\mathrm{KS}}$, where $D_{\mathrm{KS}}$ is the Kolmogorov--Smirnov distance.
For binary and categorical variables, we report TVComplement based on total variance distance,
\[
\mathrm{TVComplement}(c)
=
1 - \frac{1}{2}\sum_x \big| p_{\mathrm{real}}(x) - p_{\mathrm{synth}}(x) \big|,
\]
where $p_{\mathrm{real}}(x)$ and $p_{\mathrm{synth}}(x)$ refer to real and synthetic frequencies for the category $x$. Aggregate scores are obtained by averaging over variables.

\textbf{Pairwise dependence.}
To evaluate low-order dependence structure, we compute three complementary measures.
For continuous--continuous pairs, we report the average CorrelationSimilarity score based on
Pearson correlation. For discrete--discrete pairs, we report the average
ContingencySimilarity score based on total variation distance.
In addition, we compute symmetric uncertainty (SU), a normalized mutual information measure
\citep{yu2003feature}:
\[
\mathrm{SU}(i,j)
=
\frac{2\,\mathrm{MI}(i,j)}{H(i) + H(j)},
\]
where $\mathrm{MI}(i,j)$ denotes the mutual information between variables $i$ and $j$, and $H(i)$ and $H(j)$ are their marginal entropies. Mutual information and entropies are estimated using $k$-nearest neighbors for continuous variables \citep{kraskov2004estimating} or contingency tables with Laplace smoothing for discrete variables. We report SU similarity
\[
1 - \big|\mathrm{SU}_{\mathrm{real}}(i,j) - \mathrm{SU}_{\mathrm{synth}}(i,j) \big|
\]
averaged over all variable pairs.

\textbf{Conditional fidelity.}
We assess preservation of conditional structure by evaluating discrepancies in $p(Z \mid C)$,
where $Z$ denotes all features except the conditioning variable $C$. For a chosen conditioning
variable (e.g., treatment assignment), we partition both real and synthetic data into strata
defined by quantile bins when $C$ is continuous or by observed categories when $C$ is discrete. Within each stratum, we compute a maximum mean discrepancy (MMD$^2$) between real and synthetic
samples using a mixed kernel \citep{gretton2012kernel}, with RBF kernels for continuous variables
and Hamming kernels for discrete variables. The overall conditional fidelity score is obtained as
a stratum-size--weighted average of the per-stratum MMD$^2$ values. To contextualize the magnitude
of the discrepancy, we normalize this quantity by a within-real reference computed from a stratified split of the real dataset; values near one indicate discrepancies comparable to the intrinsic variability within the real data itself.

\textbf{Joint fidelity.}
Global fidelity is evaluated using (i) normalized energy distance on encoded representations \citep{szekely2013energy} and (ii) a classifier two-sample test (C2ST) \citep{lopez2016revisiting}.
The C2ST trains a logistic regression to distinguish real from synthetic samples and reports the ROC AUC along with the rescaled score
\[
1 - 2\,\big|\mathrm{AUC} - 0.5\big|,
\]
which attains its maximum when the two distributions are indistinguishable.

\textbf{Causal-structure diagnostics.}
Because the generator explicitly models potential outcomes and user-specified causal control
functions, we evaluate fidelity along three dimensions: treatment effects, unmeasured
confounding, and overlap.

\emph{Treatment-effect diagnostics.}
For each generated sample, the post-treatment generator produces potential outcomes
$Y'(0)$ and $Y'(1)$, yielding the induced individualized effect
\[
\tau_\theta(X') = Y'(1) - Y'(0).
\]
Given the target function $\tau(X')$, we compute the mean absolute error
\[
\mathrm{MAE}_\tau = \mathbb{E}\big[ |\tau_\theta(X') - \tau(X')| \big],
\]
the Pearson correlation $\rho(\tau_\theta(X'),\tau(X'))$ to assess heterogeneity alignment, the induced average treatment effect
$\mathrm{ATE}_\theta = \mathbb{E}[\tau_\theta(X')]$ and its deviation from the target ATE,
and the 1D Wasserstein distance between the empirical distributions of
$\tau_\theta(X')$ and $\tau(X')$. Scatter plots of $\tau_\theta(X')$ versus $\tau(X')$ and overlaid histograms provide qualitative diagnostics.

\emph{Unmeasured confounding diagnostics.}
Using both observed and flipped treatment assignments, the generator induces confounding
\[
\kappa_\theta(X',T')
=
\mathbb{E}[Y'(T')\mid X',T']
-
\mathbb{E}[Y'(1-T')\mid X',T'].
\]
Given the target $\kappa(X',T')$, we compute
\[
\mathrm{MAE}_\kappa = \mathbb{E}[|\kappa_\theta(X',T') - \kappa(X',T')|],
\]
along with group-specific MAE for treated and control units, and a Wasserstein distance between the distributions of $\kappa_\theta(X',T')$ and $\kappa(X',T')$. Scatter plots and distribution overlays provide intuitive visualization of confounding alignment.

\emph{Overlap diagnostics.}
Overlap (positivity) is evaluated using two complementary approaches. A decoder-based diagnostic computes the induced log-density ratio
\[
\Delta_\theta(X')
=
\log p_\theta(X'\mid T'=0)
-
\log p_\theta(X'\mid T'=1),
\]
which differs from the overlap control function by a sign convention. We report the mean squared
error relative to the target $\log\alpha(X')$, summary statistics of $\Delta_\theta(X')$, and the
fraction of samples within a user-specified tolerance band.
In addition, a model-agnostic, propensity-based diagnostic fits a logistic regression
$e(X')=\Pr(T'=1\mid X')$, estimates smooth propensity densities via Gaussian KDE, and computes
the overlap coefficient \citep{inman1989overlapping}
\[
\mathrm{OverlapCoeff}
=
\int_0^1 \min\big(p_{T'=0}(e),\,p_{T'=1}(e)\big)\,de.
\]
The evaluation also produces an informative plot overlaying the two density curves and shading their intersection.

Together, these diagnostics quantify how closely the generated data satisfy the user-specified
causal constraints encoded in $\tau(X)$, $\kappa(X,T)$, and $\alpha(X)$.

\textbf{Privacy.}
Privacy is defined operationally in terms of record-level disclosure risk and quantified using distance-to-closest-record (DCR)–based diagnostics, which assess whether synthetic samples are unusually close to specific real individuals. \citep{stadler2022synthetic,steier2025synthetic}. After encoding real and synthetic datasets into a common feature space, we compute for each real record the distance to its nearest synthetic neighbor and to its nearest \emph{other} real neighbor. From these distances, we report three
closely related summaries. The \emph{DCR protection fraction} is the proportion of real records for which the nearest synthetic neighbor is farther than the nearest real neighbor, with higher
values indicating stronger protection. We additionally report summary statistics of the \emph{DCR distance ratio} (synthetic–real distance divided by real–real distance), which quantifies relative closeness at the record level. Finally, we report the standardized \texttt{DCRBaselineProtection} score, which normalizes the median synthetic–real DCR by the corresponding median DCR obtained from randomly generated data; values near 1 indicate privacy protection comparable to random noise. Full details are provided in \cref{app:metrics_table}.

\subsection{CATE Estimation Methods}

Accurate estimation of CATEs is central to individualized decision-making, yet estimators from different methodological families rely on distinct modeling
assumptions and can behave very differently under confounding, limited overlap, and model misspecification. Meaningful comparison therefore requires data-generating processes with known
ground-truth effects and controllable causal violations—conditions rarely met in observational studies. \textsc{CausalMix} provides a controllable generative sandbox in which treatment assignment, outcome mechanisms, overlap, and unmeasured confounding can be systematically varied,
enabling principled stress-testing of CATE estimators across diverse causal regimes.

We evaluate representative direct CATE estimators from four major families.
\emph{Meta-learners} \citep{kunzel2019metalearners}, represented by the X-learner, construct CATE estimates by imputing individual treatment effects with weighted average of treatment effects using separate outcome models for treated and control units.
\emph{Doubly robust (DR) learners} \citep{bang2005doubly,foster2023orthogonal} combine outcome regression and propensity score modeling to form pseudo-outcomes that remain consistent if either nuisance model is correctly specified. \emph{Double machine learning (DML)} methods \citep{chernozhukov2018double} use orthogonalized estimating equations and cross-fitting to reduce bias from high-dimensional nuisance estimation. Finally, \emph{tree-based approaches} include causal forests \citep{10.1214/18-AOS1709}, which estimate heterogeneous effects via adaptive local averaging with honest sample splitting, and Bayesian causal forests (BCF) \citep{hahn2020bayesian}, which model treatment and prognostic effects separately and regularize heterogeneity through Bayesian shrinkage.

Together, these families span a broad range of modeling strategies and robustness properties, making them well suited for systematic evaluation within the \textsc{CausalMix} framework.

\section{Case Study: Heterogeneous Treatment Effects in Advanced Prostate Cancer}

\subsection{Clinical context}
Metastatic castration-resistant prostate cancer (mCRPC) is an advanced stage of prostate cancer
characterized by resistance to androgen-deprivation therapy and is associated with rapid disease
progression, poor prognosis, and limited overall survival \citep{henriquez2021current}. The most
commonly used first-line treatments for mCRPC are the novel androgen receptor--targeted therapies
abiraterone and enzalutamide, which are frequently prescribed interchangeably in routine practice
due to the absence of head-to-head phase~III randomized trials. Both therapies are associated with
clinically meaningful adverse events, and emerging observational evidence suggests that their
safety profiles may differ across patient subgroups. In particular, differential risks have been
reported among patients with pre-existing cardiovascular disease or diabetes, older adults, and
those with higher comorbidity burden \citep{schoen2023survival}.

Despite these observations, the magnitude and drivers of heterogeneity in comparative treatment
safety remain incompletely understood. Improved characterization of such heterogeneity is critical
for informing individualized treatment decisions in mCRPC. Accordingly, this case study examines
heterogeneous treatment effects in the comparative safety of abiraterone versus enzalutamide when
used as first-line therapy for mCRPC, with a focus on adverse-event risk.

\subsection{Data description}

\textbf{Data source and cohort curation.}
Empirical data were obtained from the Optum Clinformatics Data Mart, a longitudinal administrative
claims database containing information on beneficiaries with commercial and Medicare Advantage
plans. The database includes member enrollment and demographics, inpatient and outpatient medical
claims, pharmacy claims, laboratory results, and provider information. Claims from January~1,~2014
to March~31,~2024 were used in this analysis.

Patients with mCRPC were identified using a validated claims-based algorithm proposed by
Freedland et al., which incorporates diagnosis codes, prostate-specific antigen test results, drug
exposures, and procedure codes indicative of metastatic disease and castration resistance
\citep{freedland2021identification}. The algorithm was updated to include recently approved therapies. The analytic cohort comprised 4{,}098 patients who initiated first-line treatment with either enzalutamide or abiraterone following the mCRPC index date.

\textbf{Variables.}
The primary outcome was the occurrence of treatment-related adverse events, defined as hospitalizations or emergency department visits attributable to any clinically relevant adverse event during the treatment period. Treatment exposure was defined by the first filled prescription after the mCRPC index date: patients initiating abiraterone were classified as treated, while those initiating enzalutamide served as the comparison group.

Baseline covariates included 18 variables spanning binary, categorical, and continuous types. These encompassed demographic characteristics (age and race), metastatic disease sites, prior prostate cancer treatments (including taxanes and other systemic therapies), prior use of enzalutamide or abiraterone, comorbidity burden measured by the Charlson comorbidity index, history of cardiovascular disease and diabetes, other clinically relevant conditions, and concomitant
treatments with potential drug--drug interactions.

\subsection{Model training and hyperparameter selection}

The \textsc{CausalMix} generator was trained using the variational architecture described above, with hyperparameters chosen to promote stable optimization, faithful reproduction of empirical distributions, and robust control of causal targets across synthetic replications. The latent dimension was set equal to the number of target variables, while the number of mixture components in the Bayesian Gaussian mixture prior was inferred automatically. Both encoder and decoder used a
single hidden layer with 64 units, and training was performed with a batch size of 10.

Rigidity parameters controlling treatment-effect and unmeasured confounding alignment were fixed at $10^3$, while those governing overlap were set between $10^1$ and $10^2$, reflecting their different sensitivity to misspecification. Treatment-effect and confounding losses combined MSE and Smooth~L1 penalties. When control functions were low-dimensional or weakly nonlinear, MSE weights were reduced (0.2--0.4), Smooth~L1 penalties were emphasized, and variance regularization was increased to discourage collapsed or overly homogeneous synthetic effects. Variable bounds
were specified using empirical ranges from the observed data.

Training convergence and robustness were assessed through validation-loss stabilization, preservation of distributional similarity, and consistency of the controlled causal quantities across independently generated synthetic datasets. These diagnostics informed the generator validation design described next.

\subsection{CausalMix generator validation and synthetic data evaluation}

We validated the \textsc{CausalMix} generator along three complementary dimensions:
distributional fidelity, causal-structure fidelity, and record-level privacy. Our primary focus is on distributional and causal fidelity, while privacy is treated as a secondary diagnostic to ensure that increased realism and causal control do not induce excessive memorization. For each causal scenario described below, we additionally compared sampling from the proposed BGMM latent prior with a standard Gaussian prior, isolating the effect of latent multimodality on synthetic data quality.

\textbf{Causal scenario design.}
We considered three progressively more challenging causal scenarios, each defined by explicit specification of a treatment-effect function $\tau(X)$, an unmeasured confounding function $\kappa(X,T)$, and an overlap function $\alpha(X)$.

\emph{Scenario 1 (homogeneous effect, no confounding, perfect overlap).}
We specified a constant treatment effect $\tau(X)\equiv 0.1$, no unmeasured confounding $\kappa(X,T)\equiv 0$, and perfect overlap $\log\alpha(X)\equiv 0$ (equivalently, $\alpha(X)\equiv 1$). This baseline setting serves as a sanity check for verifying that \textsc{CausalMix} recovers simple causal structure while preserving the empirical data distribution.

\emph{Scenario 2 (linear heterogeneous effect, mild confounding, constant moderate overlap).}
We introduced linear treatment effect heterogeneity driven by clinically relevant covariates (pre-existing cardiovascular disease, age, and Charlson comorbidity index),
\[
\tau(X)=0.05+0.015\,\text{cvd}+0.01\,\text{age}+0.01\,\text{Charlson},
\]
with constant unmeasured confounding $\kappa(X,T)\equiv 0.02$ and moderate constant overlap $\log\alpha(X)\equiv 1$. This scenario evaluates recovery of low-dimensional heterogeneity under mild hidden bias.

\emph{Scenario 3 (nonlinear heterogeneous effect, covariate-dependent confounding and overlap).}
The most challenging setting introduced nonlinear treatment-effect heterogeneity, structured unmeasured confounding, and covariate-dependent overlap. The treatment effect was specified as a bounded nonlinear function,
\begin{equation*}
\begin{aligned}
\tau(X)
&= 0.02 + 0.05\,\tanh\!\big(
   0.4\,\text{cvd}
 + 0.2\,\text{age}
 + 0.2\,\text{Charlson}
 + 0.4\,\text{dementia}
\big),
\end{aligned}
\end{equation*}
inducing heterogeneous effects with saturation.
Unmeasured confounding depends jointly on covariates and treatment assignment,
\begin{equation*}
\begin{aligned}
\kappa(X,T)
&= 0.05 - 0.01\,(2T-1)\,\tanh\!\big(
   0.5\,\text{Charlson}
 + 0.6\,\text{cvd}
 + 0.2\,\text{age}
\big),
\end{aligned}
\end{equation*}
producing asymmetric hidden bias between treatment groups. Overlap is explicitly treatment-dependent and driven by prior exposure to abiraterone,
\[
\log\alpha(X)=2\big(2\,\text{Abiraterone\_prev}-1\big),
\]
creating regions of strong and weak overlap aligned with a clinically meaningful covariate.
This scenario stresses \textsc{CausalMix}'s ability to jointly control nonlinear effect
heterogeneity, selection bias, and treatment-dependent overlap.

\textbf{Visualization of causal fidelity.}
Quantitative causal-fidelity metrics were complemented with diagnostic visualizations. Treatment-effect fidelity was examined by comparing target and learned CATE distributions, together with scatter and joint plots that reveal alignment and dispersion across units. Unmeasured confounding was assessed using analogous comparisons. Overlap was visualized via propensity-score distributions to illustrate preservation of common support.

\textbf{Visualization of distributional fidelity.}
Distributional realism was assessed using visual diagnostics at multiple dependency scales. Marginal fidelity was examined via overlaid density plots for continuous variables and bar plots for discrete variables. Pairwise fidelity was assessed using scatter or density plots for continuous--continuous pairs, contingency-table heatmaps for discrete--discrete pairs, and discretized conditional distributions for discrete--continuous pairs. Finally, joint fidelity was visualized using 2D joint embeddings of real and synthetic samples, providing a holistic view of global alignment and revealing discrepancies not
captured by marginal or pairwise summaries.

\begin{figure*}[t]
\centering
\includegraphics[width=\textwidth]{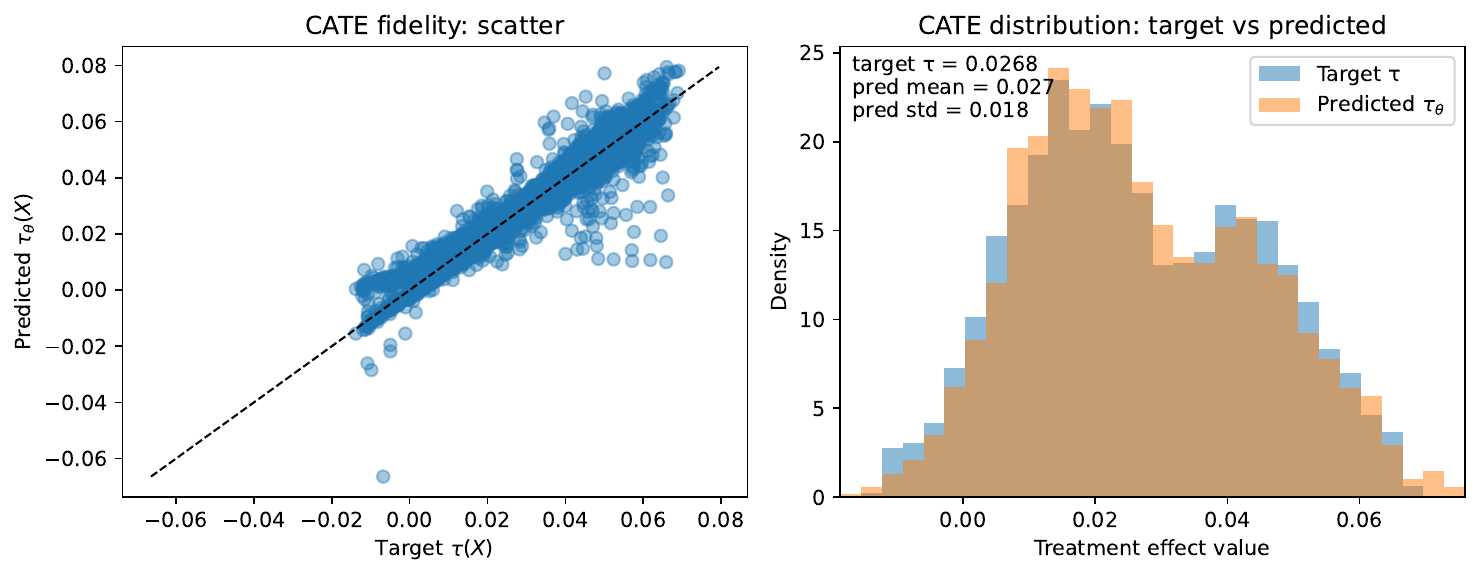}
\caption{Comparison of generated and target treatment effects of the mcrpc dataset from CausalMix with a BGMM prior in Scenario~3.}
\label{fig:m3_te}
\end{figure*}

\textbf{Results.}
Across all scenarios, synthetic data quality was evaluated along three dimensions:
distributional fidelity (Tables~\ref{tab:m1_dist}--\ref{tab:m3_dist}), causal fidelity (Tables~\ref{tab:m1_causal}--\ref{tab:m3_causal}) and privacy (Tables~\ref{tab:m1_privacy}--\ref{tab:m3_privacy}).
Diagnostic plots (\cref{fig:m3_te} and \cref{fig:m3_bias,fig:all_overlap,fig:marginals,fig:m3_exp_age,fig:m3_exp_hosp,fig:m3_exp_trt})  show that \textsc{CausalMix} achieved high distributional and causal fidelity across all dimensions under both latent priors, with performance differences emerging primarily under increasing causal complexity.

\begin{figure*}[t]
\centering
\begin{subfigure}{0.32\textwidth}
  \centering
  \includegraphics[width=\linewidth]{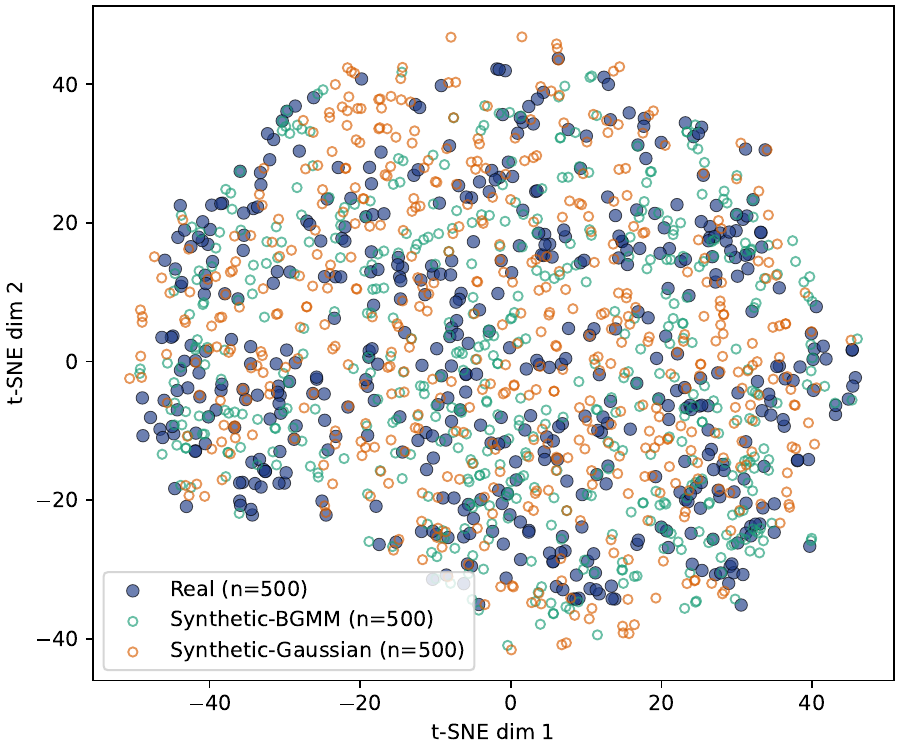}
  \caption{Scenario~1}
\end{subfigure}\hfill
\begin{subfigure}{0.32\textwidth}
  \centering
  \includegraphics[width=\linewidth]{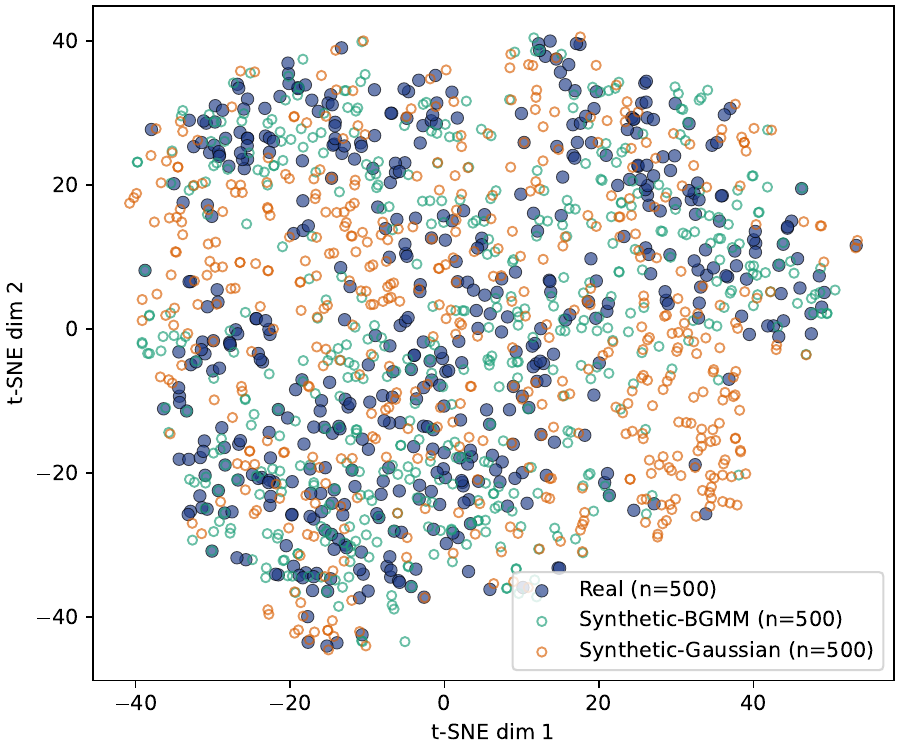}
  \caption{Scenario~2}
\end{subfigure}\hfill
\begin{subfigure}{0.32\textwidth}
  \centering
  \includegraphics[width=\linewidth]{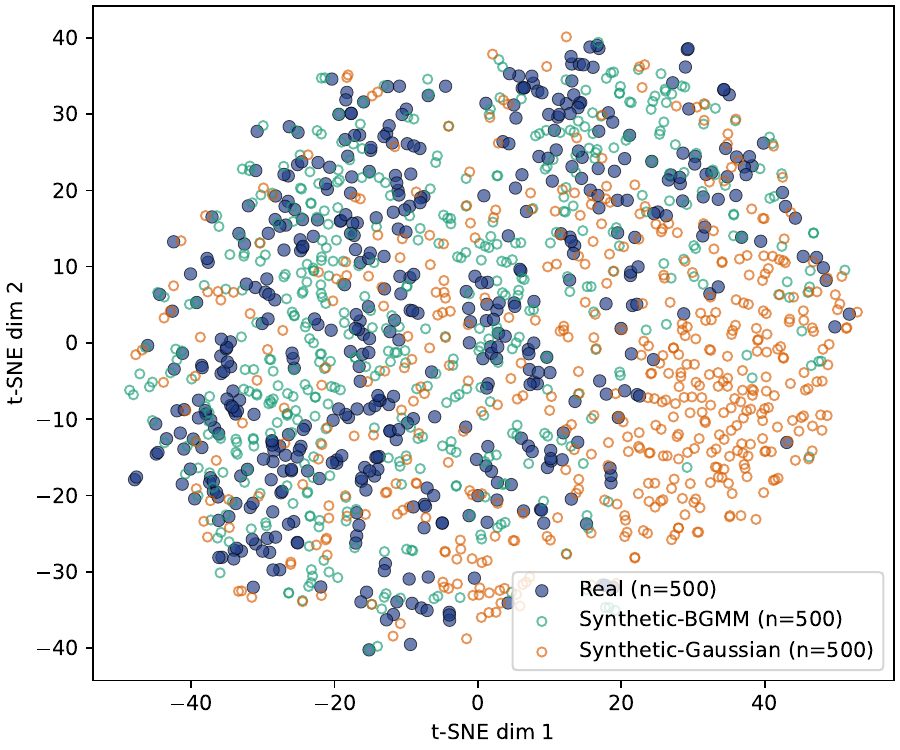}
  \caption{Scenario~3}
\end{subfigure}
\caption{2D joint embedding of synthetic data under BGMM and Gaussian priors compared with the real mCRPC dataset across causal scenarios. Scenario 1: homogeneous effect, no confounding, perfect overlap; Scenario 2: linear heterogeneous effect, mild confounding, constant overlap; Scenario 3: nonlinear heterogeneity with covariate-dependent confounding and overlap}
\label{fig:all_joint}
\end{figure*}

As illustrated by the 2D joint embeddings in Figure~\ref{fig:all_joint}, BGMM-based sampling maintained closer alignment with the real data as causal complexity increases. In Scenario~3, which combined multimodal covariate distributions, nonlinear heterogeneity, and covariate-dependent confounding and overlap, the BGMM prior substantially outperformed the Gaussian prior across marginal, pairwise, conditional, and joint distributional metrics (Table~\ref{tab:m3_dist}). In particular, BGMM yielded lower normalized Wasserstein distances for continuous marginals, higher similarity for discrete marginals and pairwise relationships, improved conditional alignment, and significantly better joint fidelity as measured by normalized energy distance and C2ST.

Both priors recovered the imposed causal structure with high accuracy. However, BGMM exhibits greater robustness in challenging regimes: in Scenario~3, it achieved higher CATE correlation and substantially more accurate decoder-level overlap reconstruction, while maintaining comparable ATE error and confounding recovery (\cref{tab:m3_causal}).

Privacy diagnostics indicate a controlled realism–privacy trade-off. The Gaussian prior consistently provided stronger record-level protection, with higher protection fractions and larger distance ratios. The BGMM prior shows moderately lower protection, most notably in Scenario~3, but maintained protection fractions well above 0.5 and median distance ratios exceeding one, indicating no systematic memorization. The reduction in protection under BGMM was concentrated in lower quantiles of the distance-ratio distribution, reflecting localized proximity rather than widespread disclosure risk (Tables~\ref{tab:m1_privacy}--\ref{tab:m3_privacy}).

Overall, these results demonstrate that \textsc{CausalMix} with a BGMM latent prior improves distributional and causal fidelity in complex clinical settings while preserving stable record-level privacy, illustrating a manageable trade-off between distributional expressiveness and disclosure
risk.

\begin{figure*}[t]
\centering
\includegraphics[width=\textwidth]{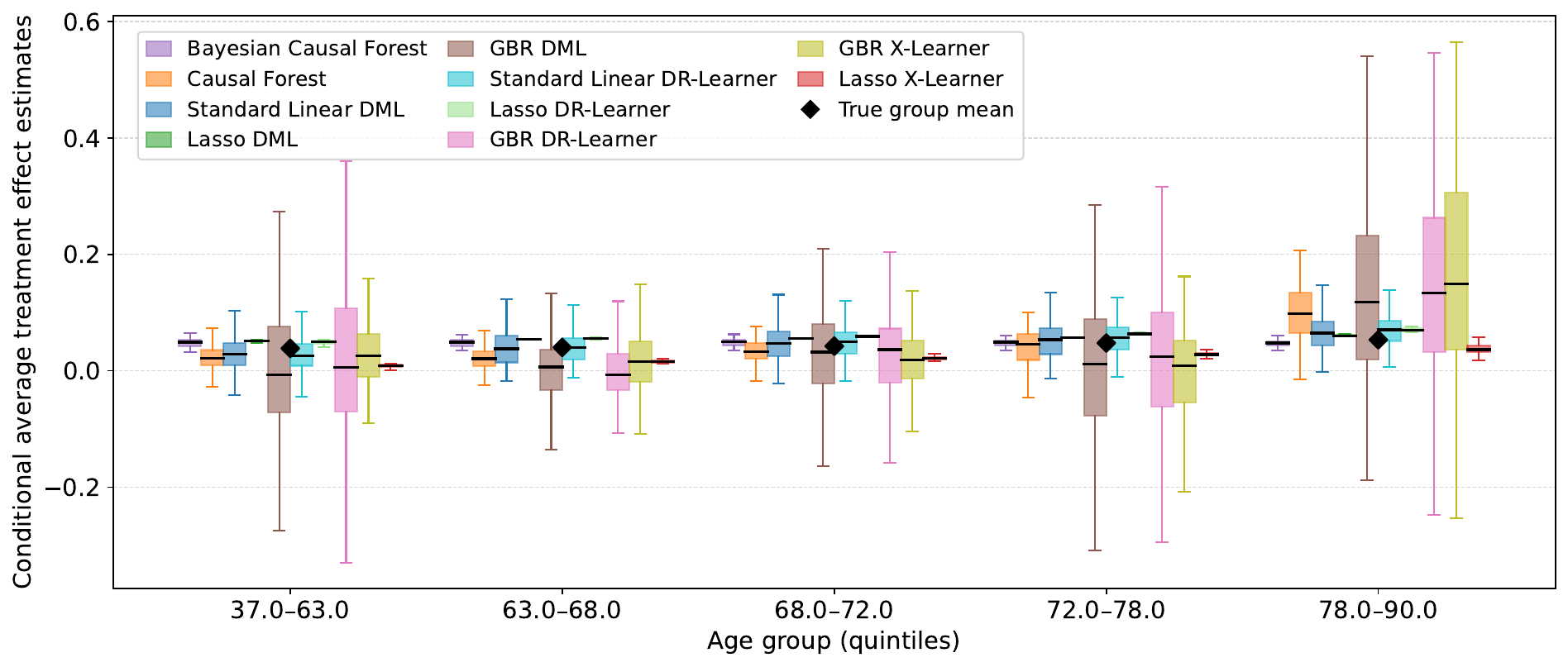}
\caption{Grouped CATE estimates by age quintile. For each age group, boxplots summarize individual-level CATE estimates from each estimator in the subgroup with Charlson = 0, no prior CVD, under a calibrated synthetic data-generating process without unmeasured confounding ($\kappa=0$). Black diamonds indicate the true group-level mean CATE.}
\label{fig:app1:cate_vs_age}
\end{figure*}

\begin{figure*}[t]
\centering
\begin{minipage}[t]{0.48\textwidth}
  \centering
  \includegraphics[width=0.8\linewidth]{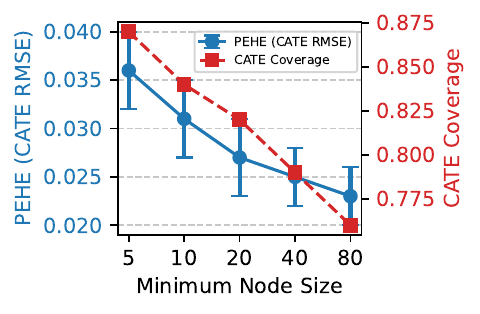}
  \captionof{figure}{PEHE vs. minimum leaf size for Causal Forests with 2{,}000 trees. Error bars indicate ±1 standard deviation across replications.}
  \label{fig:app2:pehe_coverage_min_node}
\end{minipage}\hfill
\begin{minipage}[t]{0.48\textwidth}
  \centering
  \includegraphics[width=0.8\linewidth]{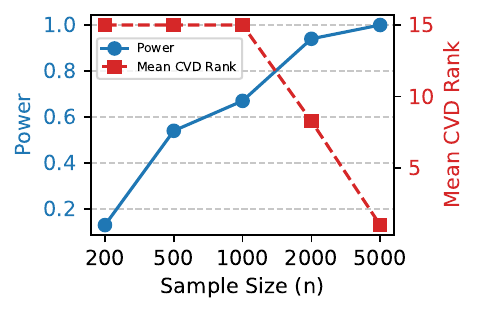}
  \captionof{figure}{Power and mean CVD rank based on the variable importance calculated from causal forests for increasing sample sizes.}
  \label{fig:app3:power_rank_n}
\end{minipage}
\end{figure*}

\subsection{Application 1: Benchmarking CATE Estimators}\label{sec:app1}

Estimating CATEs from observational data is highly sensitive to the underlying DGP, including the degree of confounding, propensity-score overlap, outcome variability, functional form (linear versus nonlinear), and the magnitude and heterogeneity of treatment effects \citep{mahajan2022empirical}. We use \textsc{CausalMix} as a controlled sandbox to benchmark CATE estimators under DGPs calibrated to match key empirical characteristics of the prostate cancer cohort. The goal is to identify estimators that are both accurate and computationally feasible for detecting heterogeneity in
comparative treatment safety.

\textbf{Design and target estimands.} We focused on the second validated causal scenario from \textsc{CausalMix}, characterized by moderate overlap ($\log\alpha(X)=1$) and linear treatment-effect heterogeneity,
\[
\tau(X)=0.05 + 0.015\cdot \texttt{cvd} + 0.01\cdot \texttt{age} + 0.01\cdot \texttt{Charlson}.
\]
The primary estimands are $\mathrm{ATE}=\mathbb{E}[\tau(X)]$ and $\text{CATE}$ $\tau(X)$. Two settings were considered: (i) no unmeasured confounding ($\kappa(X,T)=0$) and (ii) small unmeasured confounding ($\kappa(X,T)=0.02$). For each setting, we generated 50 independent synthetic datasets and retained the ground-truth average and individual treatment effects for evaluation.

\textbf{Estimators.}
We compared ten CATE estimators spanning meta-learners, orthogonalization-based methods, and Bayesian regularization approaches. Specifically, we considered:
\begin{itemize}[leftmargin=*]
    \item X-learners with gradient boosting regression (GBR) and lasso models for final-stage
    CATE estimation;
    \item DML estimators with final-stage CATE models in
    $\{\text{GBR},\ \text{lasso},\ \text{linear}\}$;
    \item DR learners with final-stage CATE models in
    $\{\text{GBR},\ \text{lasso},\ \text{linear}\}$;
    \item Tree-based methods, including causal forests and BCF.
\end{itemize}
For all non-tree estimators, gradient boosting trees were used to estimate nuisance functions (propensity scores and outcome regressions) to ensure comparability across methods. Cross-fitting was applied throughout.

\textbf{Preprocessing.}
Because linear CATE models can suffer from multicollinearity under full one-hot encoding—leading to unstable covariance estimation and unreliable asymptotic inference—we constructed two feature representations for each dataset: \textbf{dummy encoding} for X-learner, DML, and DR variants, and \textbf{full one-hot encoding} for tree-based estimators (causal forest and BCF). Continuous covariates were standardized, and binary covariates were left unchanged. Preprocessing was performed once per dataset and reused across estimators.

\textbf{Uncertainty quantification and replication.}
Asymptotic (analytic) inference was used for linear final-stage CATE models, while bootstrap inference with 100 resamples was used for lasso and nonparametric (GBR) final-stage models. Each scenario was evaluated over 50 independent replications. For computational efficiency comparisons, estimators were restricted to a single thread and wall-clock runtime was recorded.

\textbf{Evaluation metrics.}
Let $\hat{\tau}(X_i)$ denote the estimated CATE and $\tau(X_i)$ the ground-truth individual treatment effect. Performance was first computed within each synthetic dataset and then summarized across replications.

For ATE estimation, we reported the root mean squared error (RMSE) of the ATE bias across
replications, the mean and standard deviation of the ATE bias, the average estimated standard
error, and the empirical coverage probability of the ATE confidence interval.

For CATE estimation, we reported the RMSE—also known as the precision in estimating heterogeneous effects (PEHE) \citep{hill2011bayesian},
\[
\text{PEHE} = \sqrt{\frac{1}{n}\sum_{i=1}^n \big(\hat{\tau}(X_i)-\tau(X_i)\big)^2},
\]
computed within each replication and averaged across replications, along with the standard
deviation of the PEHE and the mean coverage probability of pointwise CATE intervals.

Finally, we reported the mean and standard deviation of wall-clock runtime to quantify
computational efficiency.

\textbf{Results.}
Both confounding settings share a small but nonzero true ATE of 0.056 (SD: 0.000), enabling direct
comparison of estimator accuracy and uncertainty calibration across methods.

\emph{No unmeasured confounding.}
For ATE point estimation, most estimators achieved low bias (\cref{fig:app1:ate_bias}). BCF, causal forest, and DML/DR learners attained ATE RMSEs of approximately 0.009--0.010, whereas X-learner variants exhibited substantially larger error (ATE RMSE $\approx 0.024$). Uncertainty calibration for ATE varied across methods: several estimators achieved near-nominal coverage (e.g., BCF, causal forest,
linear DML), while others attained near-perfect coverage primarily by inflating standard errors (notably GBR-based DML/DR and the linear DR learner; Table~\ref{tab:est_coverage}). Given the
limited number of Monte Carlo replications, small deviations from nominal coverage should be interpreted cautiously.

Differences across estimators were more pronounced for CATE inference than for ATE estimation. BCF and the standard linear DR learner achieved the highest mean CATE coverage (0.91--0.92), causal forest and GBR-based learners were moderately calibrated (0.81--0.87), and lasso-based DML/DR learners exhibited poor CATE coverage (0.37--0.40). This pattern indicates unreliable uncertainty quantification for heterogeneous effects despite competitive ATE RMSE (Table~\ref{tab:est_coverage}) and PEHE (\cref{fig:app1:cate_rmse}) for some methods. Figure~\ref{fig:app1:cate_vs_age} summarizes CATE estimates by age quintile within the $\text{CVD}=0,\ \text{Charlson}=0$ subgroup. Bayesian causal forest was tightly concentrated and tracked the true group means across quintiles, whereas causal forest exhibited higher variability, especially in the oldest group. Lasso-based estimators oversmoothed and attenuated fine-scale heterogeneity, consistent with their weak coverage. In contrast, gradient-boosting variants showed markedly wider spreads and heavier tails, most pronounced at older ages.

Runtime comparisons highlight practical trade-offs. Linear DML/DR learners run in approximately 2 seconds, causal forest in roughly 25 seconds, whereas BCF and boosted or lasso-based meta-learners typically require 100--270 seconds (Table~\ref{tab:est_coverage} in the appendix).

\emph{Small unmeasured confounding.}
Introducing unmeasured confounding induced systematic ATE bias across all estimators
(Figure~\ref{fig:app1:ate_bias}), with an average shift of approximately 0.02 matching the injected confounding magnitude. ATE coverage deteriorated sharply for methods with tight standard errors. Patterns in CATE coverage largely persisted: BCF remained comparatively well calibrated, whereas lasso-based DML/DR learners continued to under-cover substantially (Table~\ref{tab:est_coverage_biased} in the appendix).

\textbf{Takeaway.}
Because the prostate cancer application targets heterogeneous safety effects across patient subgroups, estimator selection should prioritize both stable CATE point estimation and calibrated
uncertainty. BCF emerges as a strong primary choice, combining accurate recovery of heterogeneous effects with high CATE coverage. DR- and DML-based approaches, particularly the standard linear DR learner, provide fast, reasonably well-calibrated baselines suitable for sensitivity analyses. Lasso-based meta-learners can capture broad CATE trends but should be treated as exploratory for inference due to poor CATE coverage. Finally, because unmeasured confounding induces systematic bias across all estimators, estimator choice alone is insufficient for validity in the real-world prostate cancer setting and should be complemented with residual confounding adjustment and sensitivity analyses.

\subsection{Application 2: Hyperparameter Optimization}

Beyond method comparison, \textsc{CausalMix} provides a principled sandbox for hyperparameter optimization of causal estimators by enabling direct evaluation against known causal ground truth under empirically calibrated data-generating processes. In this application, we used \textsc{CausalMix} to tune hyperparameters of causal forests, targeting decision-relevant accuracy in both population-level and subgroup-specific treatment effect estimation. Synthetic datasets were generated from the same calibrated regime as in \cref{sec:app1}, assuming moderate overlap, linear treatment-effect heterogeneity, and no unmeasured confounding.

\textbf{Objective.}
We treated the calibrated \textsc{CausalMix} generator $\widehat{\mathcal{G}}$ from Application~1 (without residual confounding) as a surrogate ground-truth environment and tuned causal forest hyperparameters to optimize causal performance. Given the focus on personalized and subgroup decision-making, we restricted attention to two key hyperparameters governing smoothness and stability of causal forest estimation: the minimum leaf size (\texttt{min.node.size}) and the number of trees (\texttt{num.trees}). Hyperparameter configurations were evaluated using complementary measures of estimation accuracy and uncertainty quantification.

\textbf{Experimental design.}
We considered a grid of hyperparameters defined by
\begin{align*}
\texttt{min.node.size} &\in \{5,10,20,40,80\}, \\
\texttt{num.trees} &\in \{500,1000,2000,4000\}.
\end{align*}
For each configuration, we ran 50 independent replications. In each replication, a synthetic dataset was sampled from the \textsc{CausalMix} generator $\widehat{\mathcal{G}}$, which provided both the observed data and the corresponding ground-truth ATE and CATE. A causal forest was then fitted with the specified hyperparameters, and out-of-bag predictions were used to obtain unit-level CATE estimates.

Performance was summarized at the replication level using ATE RMSE, CATE RMSE (PEHE), mean absolute CATE bias, mean CATE confidence interval width, and interval coverage for both ATE and CATE, and then averaged across replications. To characterize hyperparameter sensitivity, we jointly examined ATE accuracy and calibration alongside CATE point accuracy and uncertainty magnitude, assessing how these quantities varied across configurations. Consistent with the subgroup-oriented objective, hyperparameter selection prioritized minimizing PEHE while maintaining adequate CATE coverage, with ATE RMSE and coverage serving as population-level diagnostics.

\textbf{Results.}
Across all configurations, ATE estimation was essentially insensitive to hyperparameter choice:
ATE RMSE remained at 0.009 and ATE coverage stayed within a narrow range (0.92--0.94). In contrast,
CATE performance was driven primarily by the minimum leaf size. Increasing the minimum leaf size improved CATE point accuracy, with PEHE decreasing from approximately 0.036--0.037 (leaf size 5)
to approximately 0.023 (leaf size 80), but this improvement came at the cost of reduced CATE interval coverage, which dropped from approximately 0.87--0.91 to approximately 0.75--0.76 (see
Figure~\ref{fig:app2:pehe_coverage_min_node}). 
Consistent with this pattern, both the mean absolute CATE bias and the mean CATE confidence interval width decreased steadily as leaf size increased (Figure~\ref{fig:app2:bias_ciw}), suggesting that smaller leaf sizes induce overfitting and excess variance in this synthetic setting.
The number of trees had negligible impact on ATE and CATE accuracy or coverage once moderately large, but it dominated runtime, which increased roughly linearly (about 7\,s at 500 trees, 11\,s at 1{,}000 trees, 21\,s at 2{,}000 trees, and 42\,s at 4{,}000 trees).

Based on this trade-off, we selected a causal forest configuration with 1{,}000 or 2{,}000 trees and a
minimum leaf size of 20, which maintained strong ATE performance (ATE RMSE 0.009; ATE coverage
0.92) while balancing CATE accuracy and calibration (PEHE 0.027; mean CATE coverage 0.818) at
modest runtime (approximately 11\,s).

\textbf{Takeaway.}
These results suggest that, when using causal forests for heterogeneous safety effect estimation,
tuning the \emph{minimum leaf size} is far more consequential than substantially increasing the
number of trees. Larger leaves yield smoother CATE estimates and lower PEHEs but can produce
overly confident intervals with reduced coverage. For the mCRPC application, a practical choice
is to use approximately 1{,}000--2{,}000 trees for stability and tune the minimum leaf size
(e.g., 10--40) via sensitivity analyses, prioritizing settings that preserve clinically plausible
heterogeneity while avoiding under-covering uncertainty when reporting subgroup risk differences.

\subsection{Application 3: Power Analysis for Prospective Study Design}
\label{sec:app3_studydesign}

Detecting heterogeneous treatment effects in subgroup analyses is often limited by sample size,
particularly when effect modification is modest and covariates are correlated
\citep{imai2013estimating}. Realistic study planning therefore requires assumptions not only about
effect size, but also about the joint distribution of covariates, treatment assignment, and
outcome variability. By calibrating \textsc{CausalMix} to the empirical prostate cancer cohort, we
generate synthetic datasets that preserve these key distributional features, enabling
principled and practically relevant power analysis.

\textbf{Objective.}
In this application, we addressed a prospective study-design question:
\emph{What minimum sample size is required to reliably detect effect modification in comparative
treatment safety?} Specifically, we targeted detection of a nonzero subgroup CATE difference
associated with a pre-specified binary effect modifier,
\[
\Delta := \mathbb{E}\!\left[\tau(X)\mid \texttt{CVD}=1\right]
-
\mathbb{E}\!\left[\tau(X)\mid \texttt{CVD}=0\right],
\]
where \texttt{CVD} denotes baseline cardiovascular disease history. For clarity of interpretation,
we assumed no unmeasured confounding in this application.

\textbf{Heterogeneity setting.}
Unless otherwise noted, we adopted the same calibrated data-generating regime as in Application~2, including moderate overlap and linear treatment-effect heterogeneity. When study
assumptions warranted, heterogeneity could be varied by scaling the centered CATE component
\(h(x)=\tau(x)-\mathbb{E}[\tau(X)]\) via
\(
\tau_{\eta}(x)=\mathbb{E}[\tau(X)] + \eta\, h(x).
\)
The parameter $\eta$ rescales individual-level deviations from the ATE, allowing systematic variation of treatment-effect heterogeneity while holding the ATE fixed.
We report results for the default setting \(\eta=1\).

\textbf{Effect-modifier detection via best linear projection.}
We assessed effect modification using the best linear projection (BLP) procedure implemented in
\texttt{grf}. For each fitted causal forest, we regressed the estimated CATE on baseline
covariates using doubly robust scores,
\(\hat{\tau}(X_i) \approx \beta_0 + X_{i,S}^\top \beta\),
yielding coefficient estimates and Wald-test $p$-values with heteroskedasticity-robust (HC3)
standard errors. Inference focused on the coefficient corresponding to the pre-specified effect
modifier \texttt{CVD}. Rejection of the null hypothesis of no effect modification was based on a
two-sided test at level $\alpha=0.05$.

\textbf{Experimental design and power estimation.}
We considered candidate sample sizes
\(\mathcal{N}=\{200,500,1000,2000,5000\}\).
For each $n\in\mathcal{N}$, we generated $R=100$ independent synthetic datasets from the
calibrated \textsc{CausalMix} generator. In each replication, we fitted a causal forest using the
hyperparameters selected in Application~2 and applied the BLP procedure to test for effect
modification by \texttt{CVD}. The replication-level rejection indicator equaled one if the BLP
$p$-value for \texttt{CVD} was below $\alpha$.

Statistical power at sample size $n$ was estimated as the empirical rejection frequency across
replications. We report the resulting power curve and identify the minimum sample size
$n^\star$ achieving the target power level of 80\%.

In addition to hypothesis testing, we recorded causal forest variable-importance scores and the
corresponding ranks of the pre-specified effect modifier (\texttt{CVD}) in each replication.
These diagnostics provide complementary information on how reliably the forest identifies
\texttt{CVD} as a driver of treatment-effect heterogeneity as sample size increases.

\textbf{Results.}
For effect-modifier discovery, we fitted causal forests with 2{,}000 trees and a minimum node
size of 80 to reduce CATE noise and stabilize heterogeneity signals, without targeting calibrated
uncertainty. Figure~\ref{fig:app3:power_rank_n} summarizes empirical power (detection probability)
and the mean variable-importance rank of \texttt{CVD} across replications as functions of sample
size.

Power increased sharply with $n$: it was low at $n=200$ (0.13), rose to moderate levels at
$n=500$ (0.54) and $n=1000$ (0.67), and reached high power by $n=2000$ (0.94), approaching 1.0 at
$n=5000$. The variable-importance diagnostic exhibited a consistent pattern. When $n\le1000$,
\texttt{CVD} was not reliably prioritized (mean rank 15), whereas at $n=2000$ its mean rank
improved to 8, and by $n=5000$ it was consistently identified as the top modifier (mean rank 1).

\textbf{Takeaway.}
Under the covariate structure and heterogeneity strength considered here, reliable detection of effect modification with causal forests required on the order of $\sim$2{,}000 patients to achieve at least 90\% power. Stable prioritization of a key modifier, such as \texttt{CVD} being
ranked among the top variables, required larger samples, closer to $n=5{,}000$. For the prostate
cancer safety study, these results suggest that modifier findings from smaller cohorts should be treated as exploratory and complemented with sensitivity analyses or pooled-data strategies when robust subgroup discovery is the goal.

\section{Discussion and Conclusion}

\textbf{Summary.}
This work advances synthetic data generation as a \emph{causal inference sandbox} that jointly targets empirical realism and explicit causal control. We introduce \textsc{CausalMix}, a flexible generative framework with interpretable design-time controls for treatment effects, unmeasured confounding, and overlap, and show how mixture priors improve realism for complex,
multimodal tabular data. Beyond the generator itself, we propose a unified evaluation suite spanning distributional fidelity, causal fidelity, and privacy diagnostics, and demonstrate how validated synthetic data can guide estimator benchmarking, hyperparameter tuning, and power analysis.

\textbf{Methodological implications.}
Our results highlight a central message: causal synthetic data should be evaluated and tuned jointly for distributional realism and causal structure, with privacy risk explicitly audited. In challenging settings, BGMM-based latent sampling improves global and local distributional alignment while preserving causal fidelity. The evaluation suite provides a modular diagnostic lens—distributional metrics identify mismatches in marginals and dependence structure, causal-fidelity metrics assess realization of targeted causal quantities, and privacy
metrics quantify record-level risk—making \textsc{CausalMix} a practical testbed for causal ML development and study planning. In practice, the framework requires limited task-specific tuning: architectural choices are fixed or inferred automatically, while only a small set of interpretable causal-control parameters define the target data-generating regime.

\textbf{Insights from the case study.}
The mCRPC case study illustrates how controlled synthetic data translate generator validation into actionable methodological guidance. With known ground truth, heterogeneous-effect estimators differ along three dimensions that are often conflated in practice: ATE accuracy, uncertainty calibration, and computational cost. Notably, strong ATE accuracy does not imply reliable CATE inference—some estimators produce plausible point estimates yet substantially under-cover for CATE, limiting their suitability for subgroup-facing conclusions. Introducing unmeasured confounding induces systematic bias across all methods, reinforcing that estimator choice alone cannot ensure validity and should be complemented by unmeasured confounding diagnostics, such as negative control approaches, and sensitivity analyses.

Hyperparameter and power analyses address complementary aspects of heterogeneous treatment-effect analysis. Hyperparameter tuning focuses on estimation quality: for causal forests, the minimum leaf size governs the bias–variance trade-off in CATE estimation, with larger leaves producing smoother and more stable effect estimates at the cost of local adaptivity (finer
heterogeneity), while the number of trees primarily affects computational stability once sufficiently large. In contrast, power analysis targets effect-modifier identification and study design. Our results show that detecting the presence of treatment-effect heterogeneity requires substantially fewer samples than reliably identifying and prioritizing specific modifiers. While moderate sample sizes may suffice to reject homogeneity, robust subgroup discovery demands larger cohorts, highlighting the distinction between accurate estimation of heterogeneous effects and actionable identification of effect modifiers.

Together, these components make \textsc{CausalMix} suitable both as a testbed for causal machine
learning development and as a decision-support tool for observational studies where estimator choice and study design depend on heterogeneous effects.

\textbf{Limitations and future work.}
As with any learned generator, \textsc{CausalMix} extrapolates according to the structural assumptions of its neural architecture in regions of the covariate space that are sparsely supported by data; as a result, strong performance on aggregate distributional and causal-fidelity metrics does not guarantee faithful reproduction of rare events, extreme covariate combinations, or high-order dependencies that are weakly represented in the empirical data.

Moreover, causal control relies on user-specified functional forms. While low-dimensional, interpretable specifications improve stability and transparency, they may fail to capture complex interaction structure when true treatment-effect heterogeneity is high-dimensional. This limitation reflects a broader tension between distributional realism and causal controllability:
not all causal targets can be simultaneously enforced with high fidelity, particularly when they conflict with the empirical structure of the data. Developing principled strategies for specifying, validating, and stress-testing causal function therefore remains an important direction for future work.

A practical limitation concerns training data requirements. As a flexible neural generator, \textsc{CausalMix} requires sufficient sample size to reliably capture multimodal covariate structure and heterogeneous treatment effects while stabilizing causal-control constraints.
In smaller datasets, overfitting or instability may arise, particularly as dimensionality and effect complexity increase. A systematic characterization of data-size thresholds for stable causal synthetic generation remains an open direction for future work.

Finally, \textsc{CausalMix} currently targets static treatment settings with complete covariate information. Extending the framework to dynamic treatment regimes and explicit missing-data mechanisms would improve realism for longitudinal healthcare data and enable stress-testing of causal estimators under more realistic data availability constraints.

\bibliography{references}
\bibliographystyle{unsrt}

\medskip
\appendix
\crefalias{section}{appendix}
\crefname{appendix}{Appendix}{Appendices}
\Crefname{appendix}{Appendix}{Appendices}
\crefname{section}{Appendix}{Appendices} 
\Crefname{section}{Appendix}{Appendices} 
\onecolumn
\section{Computation of Log-Density Ratios for Mixed Variable Types}
\label{app:llr}
This appendix provides the explicit formulas used to compute the conditional log-likelihoods
\[
\log p_\theta(X \mid T'=t), \qquad t \in \{0,1\},
\]
for mixed-type covariates in the pre-treatment generator. These terms are required for evaluating
the induced overlap function
\[
\log \alpha_\theta(X)
=
\log p_\theta(X \mid T'=1)
-
\log p_\theta(X \mid T'=0),
\]
which enters the overlap-matching penalty in the training objective.

We assume a conditional independence structure within the decoder:
\[
p_\theta(X \mid z, T'=t)
=
\prod_{i=1}^V p_\theta(X_i \mid z, T'=t),
\]
so that the total log-likelihood decomposes as
\[
\log p_\theta(X \mid z, T'=t)
=
\sum_{i=1}^V \log p_\theta(X_i \mid z, T'=t).
\]
The formulas below summarize the per-variable contribution for binary, categorical, and continuous covariates.

\subsection*{A.1 Binary Variables}

For a binary variable \(X_i \in \{0,1\}\), the decoder outputs a logit \(s_i\), with
\[
p_i = \sigma(s_i) = \frac{1}{1 + e^{-s_i}}.
\]
The Bernoulli log-likelihood contribution is
\[
\log p_\theta(X_i \mid z, T'=t)
=
X_i \log p_i + (1-X_i)\log(1-p_i).
\]
This is equivalent to the negative of \texttt{binary\_cross\_entropy\_with\_logits} computed
per-sample.

\subsection*{A.2 Categorical Variables}

For a categorical variable \(X_i \in \{0,\dots,K-1\}\), the decoder outputs logits
\(s_i \in \mathbb{R}^K\). The softmax probabilities are
\[
p_{i,c}
=
\frac{\exp(s_{i,c})}{\sum_{k=1}^K \exp(s_{i,k})}.
\]
The log-likelihood contribution for the observed class \(X_i\) is
\[
\log p_\theta(X_i \mid z, T'=t)
=
\log p_{i,X_i}
=
s_{i,X_i} - \log\!\left(\sum_{k=1}^K e^{s_{i,k}}\right),
\]
which corresponds to the negative per-sample \texttt{cross\_entropy} loss.

\subsection*{A.3 Continuous Variables}

For a continuous variable \(X_i \in \mathbb{R}\), the decoder outputs a mean \(\mu_i\) and
a bounded log-variance \(\log \sigma_i^2\). The model assumes
\[
X_i \mid z, T'=t \sim \mathcal{N}(\mu_i, \sigma_i^2),
\]
and thus the Gaussian log-density is
\[
\log p_\theta(X_i \mid z, T'=t)
=
-\frac{1}{2}
\left[
\log(2\pi)
+
\log \sigma_i^2
+
\frac{(X_i - \mu_i)^2}{\sigma_i^2}
\right].
\]

\subsection*{A.4 Log-Density Ratio Computation}

Summing the feature-wise terms yields the conditional log-likelihoods
\[
\log p_\theta(X \mid T'=t)
=
\sum_{i=1}^V \log p_\theta(X_i \mid z, T'=t).
\]
The induced log-density ratio used in the overlap penalty is then
\[
\log \alpha_\theta(X)
=
\log p_\theta(X \mid T'=1)
-
\log p_\theta(X \mid T'=0).
\]

The overlap-control term in the main objective penalizes squared deviation from the
user-specified target:
\[
\mathcal{L}_{\alpha}(\theta)
=
\mathbb{E}_{X}
\left[
\big(\log\alpha_\theta(X) - \log\alpha(X)\big)^2
\right].
\]
This mechanism enables fine-grained control over positivity and covariate overlap in the synthetic data distribution.

\clearpage
\section{Summary of Evaluation Metrics}
\label{app:metrics_table}
\renewcommand{\thefigure}{B.\arabic{figure}}
\setcounter{figure}{0}
\renewcommand{\thetable}{B.\arabic{table}}
\setcounter{table}{0}
\begin{table}[!htbp]
\caption{\textbf{Summary of evaluation metrics across three axes: distributional fidelity, causal-structure fidelity, and privacy.}}
\label{tab:metric_summary}
\centering
\footnotesize
\renewcommand{\arraystretch}{1.05}
\setlength{\tabcolsep}{3pt}
\begin{tabularx}{\textwidth}{
  >{\raggedright\arraybackslash}p{2.4cm} 
  >{\raggedright\arraybackslash}p{2.8cm} 
  >{\raggedright\arraybackslash}p{3.0cm} 
  >{\raggedright\arraybackslash}X 
}
\toprule
\textbf{Axis} & \textbf{Category (Var Types)} & \textbf{Metric} & \textbf{Description / Formula} \\
\midrule

\multirow[t]{9}{=}{\textbf{Distributional\\Fidelity}}
& Marginal (cont.) & Normalized Wasserstein
& $\dfrac{W_1\!\big(F_{\mathrm{real}}^{(c)},F_{\mathrm{synth}}^{(c)}\big)}{\mathrm{sd}_{\mathrm{real}}^{(c)}+\varepsilon}$. \\

& Marginal (cont.) & KSComplement (\texttt{sdmetrics})
& $1-D_{\mathrm{KS}}$. \\

& Marginal (disc.) & TVComplement (\texttt{sdmetrics})
& $1-\dfrac{1}{2}\sum_x \big|p_{\mathrm{real}}(x)-p_{\mathrm{synth}}(x)\big|$. \\

& Pairwise (all vars) & SU Similarity
& $1-\big|\mathrm{SU}_{\mathrm{real}}(i,j)-\mathrm{SU}_{\mathrm{synth}}(i,j)\big|$. \\

& Pairwise (cont.--cont.) & CorrelationSimilarity (\texttt{sdmetrics})
& Similarity of Pearson correlations over cont. pairs. \\

& Pairwise (disc.--disc.) & ContingencySimilarity (\texttt{sdmetrics})
& Similarity of disc. contingency structure via total variation. \\

& Conditional ($Z\mid C$) & Weighted / Normalized MMD$^2$
& Stratum-size weighted mixed-kernel $\mathrm{MMD}^2(Z\mid C)$; normalized by within-real baseline. \\

& Joint (all vars) & Energy Dist.
& Normalized energy distance. \\

& Joint (all vars) & C2ST
& $1-2|\mathrm{AUC}-0.5|$. \\

\midrule

\multirow[t]{9}{=}{\textbf{Causal-Structure\\Fidelity}}
& Treatment Effect & ATE Error
& $|\mathrm{ATE}_\theta-\mathrm{ATE}_{\mathrm{target}}|$. \\

& Treatment Effect & CATE MAE
& $\mathbb{E}\!\big[\,|\tau_\theta(X')-\tau(X')|\,\big]$. \\

& Treatment Effect & CATE Correlation
& Pearson correlation $\rho(\tau_\theta,\tau)$. \\

& Treatment Effect & Wasserstein Dist.
& $W_1$ between empirical distributions of $\tau_\theta(X')$ and $\tau(X')$. \\

& Confounding & MAE (overall / by group)
& $\mathbb{E}[|\kappa_\theta(X',T')-\kappa(X',T')|]$; by treatment arm. \\

& Confounding & Wasserstein Dist.
& $W_1$ between empirical confounding distributions. \\

& Overlap & MSE
& $\mathbb{E}\!\big[(\log\alpha_\theta(X')-\log\alpha(X'))^2\big]$. \\

& Overlap & Overlap Coefficient
& $\int_0^1 \min\big(p_{0}(e),p_{1}(e)\big)\,de$. \\

& Overlap & Propensity AUC
& AUC of logistic regression predicting $T'$ from $X'$. \\

\midrule

\multirow[t]{3}{=}{\textbf{Privacy}}
& DCR & Protection Fraction
& $\frac{1}{n}\sum_{i=1}^n \mathbf{1}\{d_{\mathrm{syn}}(i)>d_{\mathrm{real}}(i)\}$. \\

& DCR & Distance Ratio
& $\rho_i=\dfrac{d_{\mathrm{syn}}(i)}{d_{\mathrm{real}}(i)+\varepsilon}$. \\

& DCR & DCRBaselineProtection (\texttt{sdmetrics})
& $\min\!\left(1,\ \widetilde{d}_{\mathrm{synth}}/\widetilde{d}_{\mathrm{rand}}\right)$. \\

\bottomrule
\end{tabularx}


\vspace{2pt}
\begin{minipage}{\textwidth}
\footnotesize
\emph{Abbreviations and notation:
cont.\ = continuous; disc.\ = discrete (binary or categorical); TE = treatment effect; dist.\ = distance;
$C$ = conditioning variable; $Z$ = all variables except $C$.
For continuous feature $c$: $F_{\mathrm{real}}^{(c)},F_{\mathrm{synth}}^{(c)}$ = empirical CDFs;
$W_1$ = 1-Wasserstein distance; $D_{\mathrm{KS}}$ = Kolmogorov--Smirnov statistic.
For discrete $x$: $p_{\mathrm{real}}(x),p_{\mathrm{synth}}(x)$ = empirical PMFs.
$\mathrm{SU}_{\mathrm{real}}(i,j),\mathrm{SU}_{\mathrm{synth}}(i,j)$ = symmetric uncertainty (real vs.\ synthetic).
$\mathrm{MMD}^2$ = maximum mean discrepancy.
$(X',T')$ = synthetic samples;
$\mathrm{ATE}_{\mathrm{target}}=\mathbb{E}[\tau(X')]$,
$\mathrm{ATE}_\theta=\mathbb{E}[\tau_\theta(X')]$.
$e=\Pr(T'=1\mid X')$; $p_0(e),p_1(e)$ = group-specific densities.
DCR = distance to closest record (encoded space);
$d_{\mathrm{syn}}(i),d_{\mathrm{real}}(i)$ = NN distances to synthetic and to another real record;
$\widetilde d_{\mathrm{synth}},\widetilde d_{\mathrm{rand}}$ = median DCRs (synthetic, random).
Metrics marked \texttt{sdmetrics} use the \texttt{sdmetrics} library.}

\end{minipage}
\end{table}

\newpage
\section{Privacy Evaluation}

The distance-to-closest-record (DCR) is a nearest-neighbor–based criterion that
quantifies how well synthetic data avoid reproducing or closely approximating real
individuals \citep{stadler2022synthetic,steier2025synthetic}. Let $\mathcal{R} = \{r_i\}_{i=1}^n$ and $\mathcal{S} = \{s_j\}_{j=1}^m$ denote the
encoded real and synthetic datasets, respectively, after standardizing continuous variables
and one-hot encoding discrete variables.

For each real record $r_i$, we compute two distances:
\[
d_{\mathrm{syn}}(i)
=
\min_{s_j \in \mathcal{S}}
\big\| r_i - s_j \big\|,
\qquad
d_{\mathrm{real}}(i)
=
\min_{\substack{r_k \in \mathcal{R} \\ k \neq i}}
\big\| r_i - r_k \big\|.
\]

The protection criterion for $r_i$ is defined as:
\[
\mathbf{1}_{\mathrm{protected}}(i)
=
\mathbf{1}\!\left\{
d_{\mathrm{syn}}(i) > d_{\mathrm{real}}(i)
\right\},
\]
that is, a real record is considered protected if its closest synthetic neighbor is farther away than its closest \emph{other} real neighbor. The DCR protection fraction is therefore
\[
\mathrm{DCR\;Protection}
=
\frac{1}{n}
\sum_{i=1}^n
\mathbf{1}_{\mathrm{protected}}(i),
\]
with larger values indicating stronger privacy protection.

To provide finer resolution, we compute the distance ratio:
\[
\rho(i)
=
\frac{d_{\mathrm{syn}}(i)}
     {d_{\mathrm{real}}(i) + \varepsilon},
\]
and report summary statistics including its mean, median, and selected quantiles (e.g.,
$5$th, $50$th, $95$th percentiles). Ratios significantly greater than one suggest that the
synthetic dataset is well separated from the real data in feature space.

We additionally report the \texttt{DCRBaselineProtection} score from the
\texttt{sdmetrics} library. It normalized median DCR of synthetic data by the median DCR of random data (clipped to $[0,1]$), with values near 1 indicating privacy comparable to random noise.

\[
\mathrm{DCRBaselineProtection}
=
\min\!\left(
1,\;
\frac{\operatorname{median}_{x \in \mathcal{D}_{\mathrm{synth}}}
\; d(x,\mathcal{D}_{\mathrm{real}})}
{\operatorname{median}_{r \in \mathcal{D}_{\mathrm{rand}}}
\; d(r,\mathcal{D}_{\mathrm{real}})}
\right)
\]

\newpage
\FloatBarrier 
\section{Additional Results}
\renewcommand{\thefigure}{D.\arabic{figure}}
\setcounter{figure}{0}
\renewcommand{\thetable}{D.\arabic{table}}
\setcounter{table}{0}
\label{app:results}
\begin{table*}[h]
\centering
\caption{Distributional fidelity using BGMM and standard Gaussian priors in Scenario~1.}
\label{tab:m1_dist}
\adjustbox{max width=\textwidth, center}{
\begin{threeparttable}
\begin{tabular}{llrrr}
\toprule
                  Category &                            Metric &                        Direction &          BGMM &      Gaussian \\
\midrule
          Marginal (cont.) &     Normalized Wasserstein (mean) &                         ↓ better & 0.071 (0.001) & 0.336 (0.010) \\
          Marginal (cont.) &               KSComplement (mean) &                         ↑ better & 0.939 (0.001) & 0.853 (0.006) \\
          Marginal (disc.) &               TVComplement (mean) &                         ↑ better & 0.973 (0.001) & 0.939 (0.001) \\
     Pairwise (cont–cont.) &             CorrelationSimilarity &                         ↑ better & 0.987 (0.002) & 0.915 (0.005) \\
       Pairwise (all vars) &              SU similarity (mean) &                         ↑ better & 0.994 (0.002) & 0.992 (0.000) \\
     Pairwise (disc–disc.) &      ContingencySimilarity (mean) &                         ↑ better & 0.953 (0.001) & 0.899 (0.002) \\
Conditional (all except C)$^{a}$ &                  Weighted MMD$^2$ &                         ↓ better & 0.004 (0.001) & 0.007 (0.002) \\
Conditional (all except C)$^{a}$ & Normalized MMD$^2$ ratio vs real &                         ↓ better & 1.533 (0.154) & 2.531 (0.601) \\
          Joint (all vars) &        Normalized Energy Distance &                         ↓ better & 0.004 (0.001) & 0.025 (0.001) \\
          Joint (all vars) &             C2ST (AUC complement) &                         ↑ better & 0.748 (0.009) & 0.551 (0.013) \\
\bottomrule
\end{tabular}

\begin{tablenotes}[flushleft]
\footnotesize
\item[a] Conditional variable (C) is exposure.
\end{tablenotes}
\end{threeparttable}
}
\end{table*}

\begin{table*}[h]
\centering
\caption{Distributional fidelity using BGMM and standard Gaussian priors in Scenario~2.}
\label{tab:m2_dist}
\adjustbox{max width=\textwidth,center}{
\begin{threeparttable}
\begin{tabular}{llrrr}
\toprule
                  Category &                            Metric &                        Direction &          BGMM &      Gaussian \\
\midrule
          Marginal (cont.) &     Normalized Wasserstein (mean) &                         ↓ better & 0.111 (0.004) & 0.360 (0.011) \\
          Marginal (cont.) &               KSComplement (mean) &                         ↑ better & 0.957 (0.001) & 0.858 (0.004) \\
          Marginal (disc.) &               TVComplement (mean) &                         ↑ better & 0.978 (0.001) & 0.928 (0.002) \\
     Pairwise (cont–cont.) &             CorrelationSimilarity &                         ↑ better & 0.985 (0.004) & 0.950 (0.007) \\
       Pairwise (all vars) &              SU similarity (mean) &                         ↑ better & 0.993 (0.002) & 0.990 (0.000) \\
     Pairwise (disc–disc.) &      ContingencySimilarity (mean) &                         ↑ better & 0.956 (0.001) & 0.881 (0.003) \\
Conditional (all except C)$^{a}$ &            Weighted MMD$^2$ &                         ↓ better & 0.006 (0.001) & 0.016 (0.001) \\
Conditional (all except C)$^{a}$ &  Normalized MMD$^2$ ratio vs real &                   ↓ better & 2.244 (0.446) & 5.981 (0.358) \\
          Joint (all vars) &        Normalized Energy Distance &                         ↓ better & 0.005 (0.000) & 0.033 (0.002) \\
          Joint (all vars) &             C2ST (AUC complement) &                         ↑ better & 0.699 (0.016) & 0.510 (0.016) \\
\bottomrule
\end{tabular}

\begin{tablenotes}[flushleft]
\footnotesize
\item[a] Conditional variable (C) is exposure.
\end{tablenotes}
\end{threeparttable}
}
\end{table*}

\begin{table*}[h]
\centering
\caption{Distributional fidelity using BGMM and standard Gaussian priors in Scenario~3.}
\label{tab:m3_dist}
\adjustbox{max width=\textwidth,center}{
\begin{threeparttable}
\begin{tabular}{llrrr}
\toprule
                  Category &                            Metric &                        Direction &          BGMM &       Gaussian \\
\midrule
          Marginal (cont.) &     Normalized Wasserstein (mean) &                         ↓ better & 0.075 (0.001) &  0.679 (0.008) \\
          Marginal (cont.) &               KSComplement (mean) &                         ↑ better & 0.954 (0.001) &  0.720 (0.005) \\
          Marginal (disc.) &               TVComplement (mean) &                         ↑ better & 0.961 (0.001) &  0.898 (0.002) \\
     Pairwise (cont–cont.) &             CorrelationSimilarity &                         ↑ better & 0.984 (0.002) &  0.883 (0.009) \\
       Pairwise (all vars) &              SU similarity (mean) &                         ↑ better & 0.991 (0.002) &  0.990 (0.000) \\
     Pairwise (disc–disc.) &      ContingencySimilarity (mean) &                         ↑ better & 0.928 (0.001) &  0.841 (0.003) \\
Conditional (all except C)$^{a}$ &            Weighted MMD$^2$ &                         ↓ better & 0.023 (0.002) &  0.029 (0.002) \\
Conditional (all except C)$^{a}$ & Normalized MMD$^2$ ratio vs real &                    ↓ better & 8.563 (0.604) & 10.863 (0.724) \\
          Joint (all vars) &        Normalized Energy Distance &                         ↓ better & 0.011 (0.000) &  0.080 (0.002) \\
          Joint (all vars) &             C2ST (AUC complement) &                         ↑ better & 0.625 (0.008) &  0.333 (0.010) \\
\bottomrule
\end{tabular}

\begin{tablenotes}[flushleft]
\footnotesize
\item[a] Conditional variable (C) is exposure.
\end{tablenotes}
\end{threeparttable}
}
\end{table*}

\begin{figure}[h]
\centering
\includegraphics[width=\textwidth]{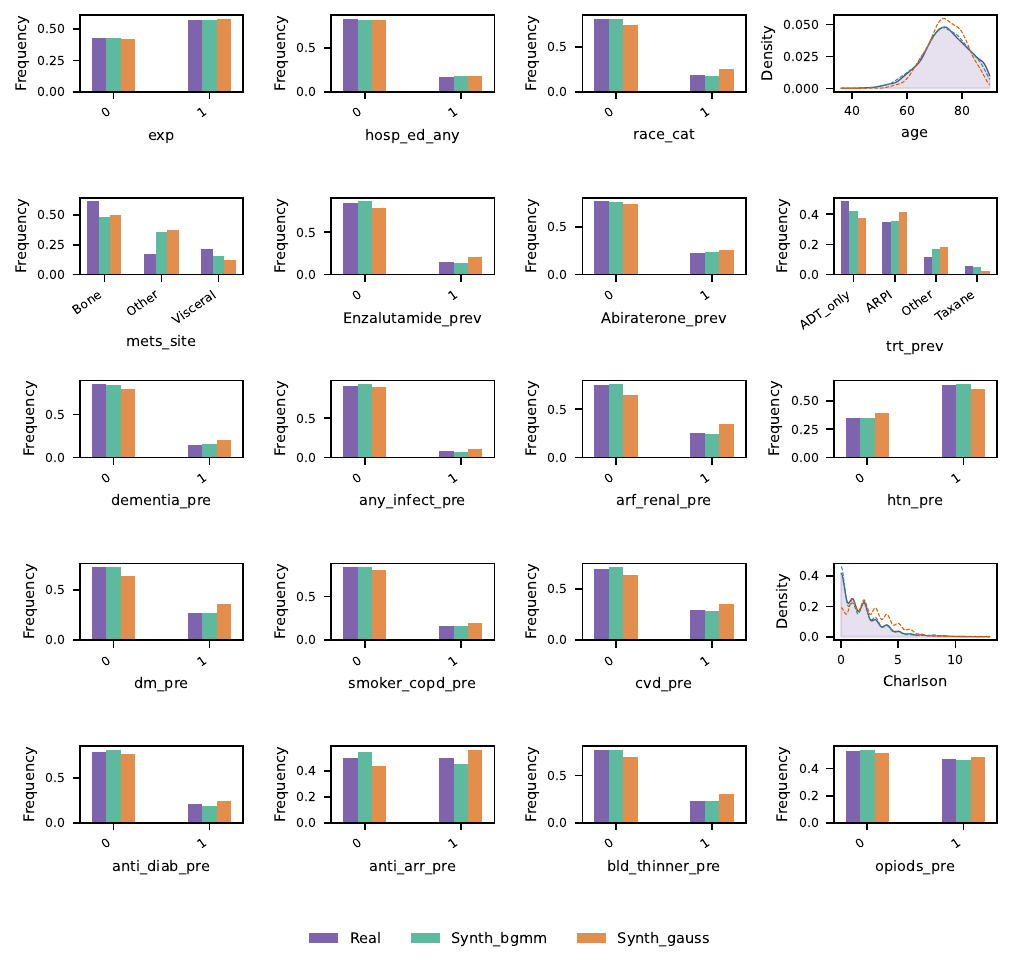}
\caption{Marginal distributions of all variables in the synthetic data under BGMM and Gaussian priors compared with the real mCRPC dataset (Scenario~3).}
\label{fig:marginals}
\end{figure}

\begin{figure*}[h]
\centering
\includegraphics[width=\textwidth]{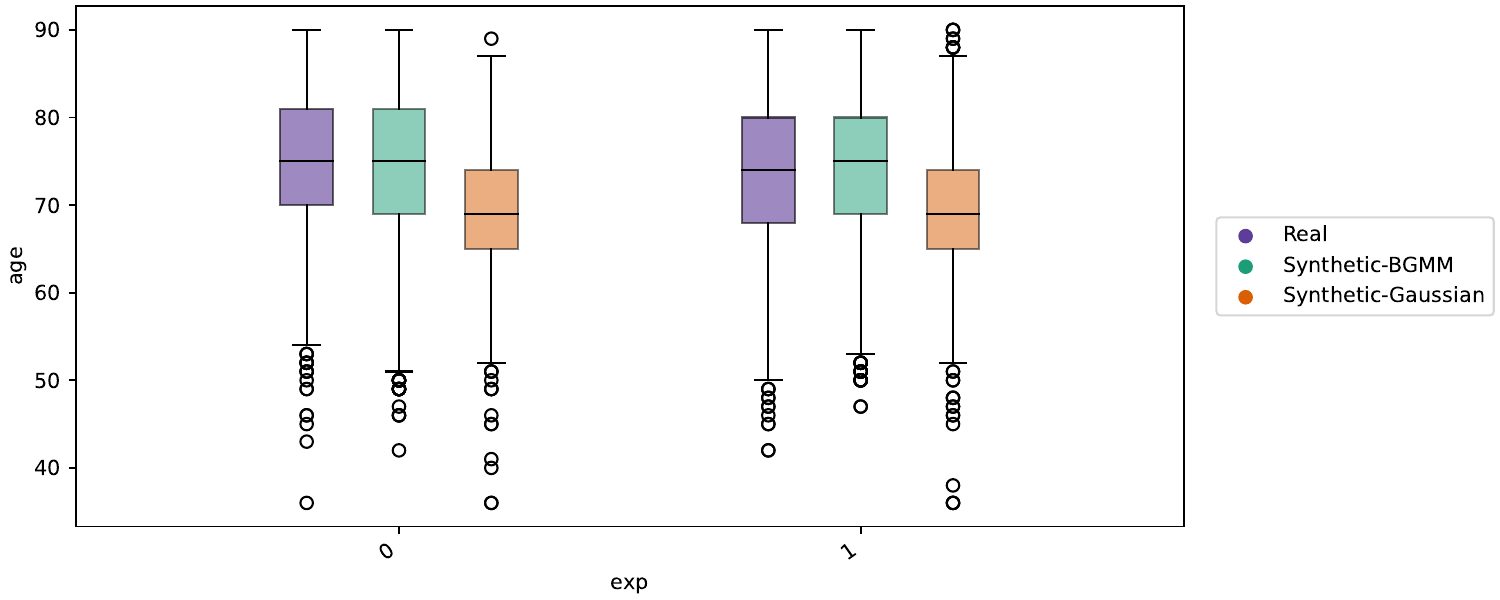}
\caption{Pairwise distributions of exposure and age in the synthetic data under BGMM and Gaussian priors compared with the real mCRPC dataset (Scenario~3).}
\label{fig:m3_exp_age}
\end{figure*}

\begin{figure*}[h]
\centering
\includegraphics[width=\textwidth]{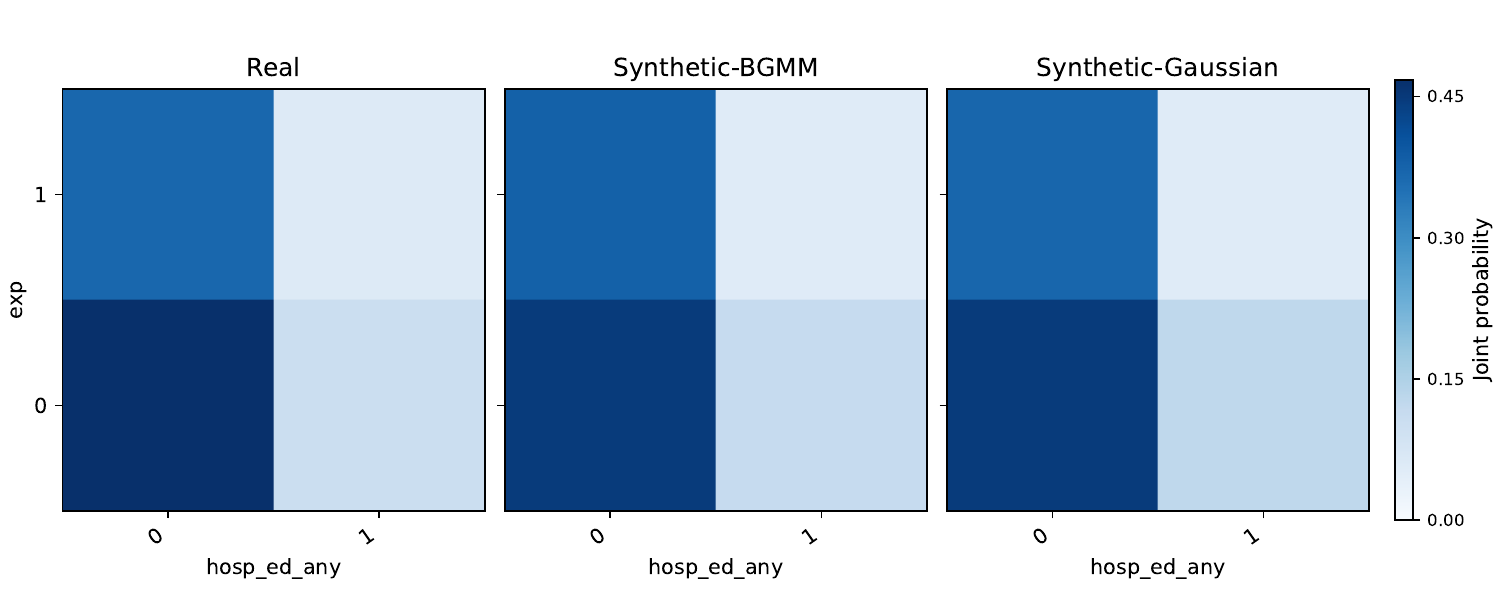}
\caption{Pairwise distributions of exposure and outcome in the synthetic data under BGMM and Gaussian priors compared with the real mCRPC dataset (Scenario~3).}
\label{fig:m3_exp_hosp}
\end{figure*}

\begin{figure*}[h]
\centering
\includegraphics[width=\textwidth]{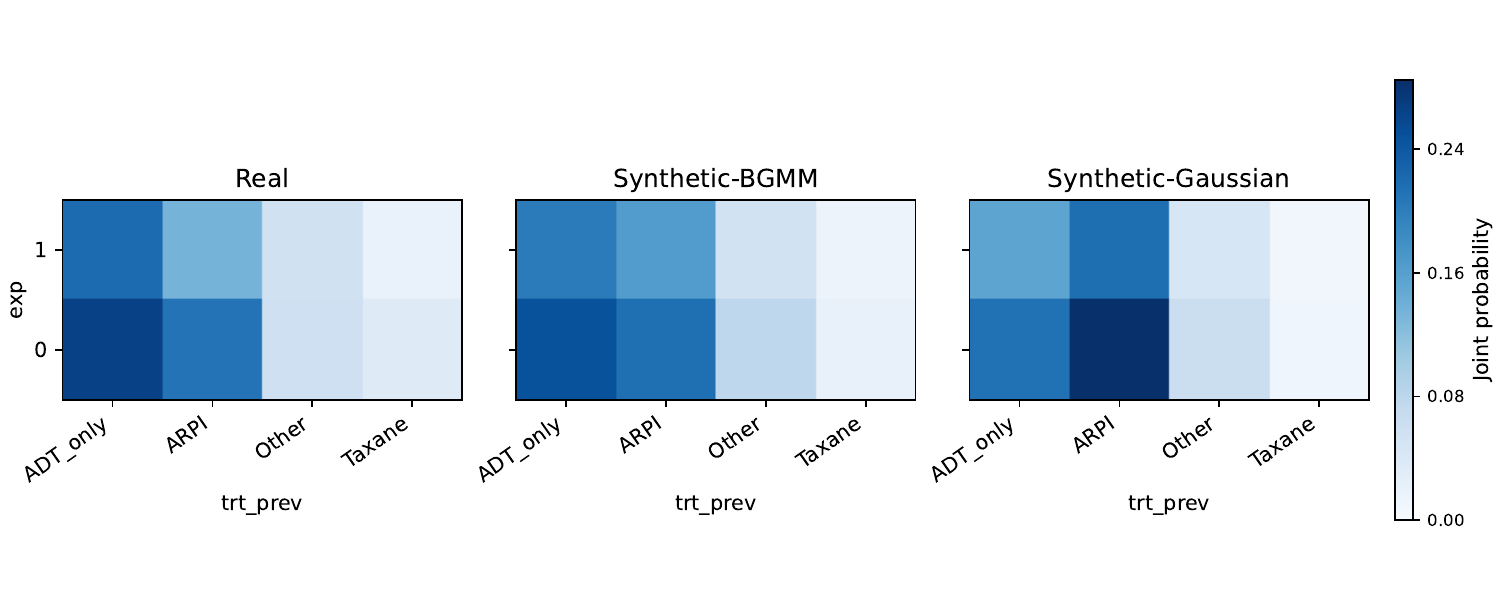}
\caption{Pairwise distributions of exposure and treatment history in the synthetic data under BGMM and Gaussian priors compared with the real mCRPC dataset (Scenario~3).}
\label{fig:m3_exp_trt}
\end{figure*}

\begin{table*}[h]
\centering
\caption{Causal-structure fidelity of synthetic data using BGMM and standard Gaussian priors in Scenario~1.}
\label{tab:m1_causal}
\adjustbox{max width=\textwidth, center}{
\begin{threeparttable}
\begin{tabular}{lllrr}
\toprule
            Category &                        Metric &                                Direction &          BGMM &      Gaussian \\
\midrule
    Treatment Effect &                  CATE/ITE MAE &                                 ↓ better & 0.003 (0.000) & 0.004 (0.000) \\
    Treatment Effect &              CATE Correlation &                                 ↑ better &           NA  &            NA \\
    Treatment Effect &                     ATE Error &                                 ↓ better & 0.000 (0.000) & 0.002 (0.000) \\
    Treatment Effect & TE Distribution Distance (W1) &                                 ↓ better & 0.003 (0.000) & 0.004 (0.000) \\
         Confounding &               Confounding MAE &                                 ↓ better & 0.000 (0.000) & 0.000 (0.000) \\
         Confounding &          Group-wise MAE (T=0) &                                 ↓ better & 0.000 (0.000) & 0.000 (0.000) \\
         Confounding &          Group-wise MAE (T=1) &                                 ↓ better & 0.000 (0.000) & 0.000 (0.000) \\
         Confounding &        Confounding Dist. (W1) &                                 ↓ better & 0.000 (0.000) & 0.000 (0.000) \\
   Overlap (decoder) &                           MSE &                                 ↓ better & 0.000 (0.000) & 0.000 (0.000) \\
   Overlap (decoder) &     Fraction within tolerance &                                 ↑ better & 1.000 (0.000) & 1.000 (0.000) \\
Overlap (propensity) &          Propensity AUC$^{a}$ &                                 NA       & 0.539 (0.007) & 0.536 (0.008) \\
Overlap (propensity) & Histogram overlap coefficient &                                 ↑ better & 0.940 (0.017) & 0.947 (0.015) \\
Overlap (propensity) &       Common support fraction &                                 ↑ better & 1.000 (0.000) & 1.000 (0.000) \\
Overlap (propensity) & Common support fraction (T=0) &                                 ↑ better & 1.000 (0.000) & 1.000 (0.000) \\
Overlap (propensity) & Common support fraction (T=1) &                                 ↑ better & 1.000 (0.000) & 1.000 (0.000) \\
\bottomrule
\end{tabular}

\begin{tablenotes}[flushleft]
\footnotesize
\item[a] AUC closer to 0.5 indicates stronger overlap.
\end{tablenotes}
\end{threeparttable}
}
\end{table*}

\begin{table*}[h]
\centering
\caption{Causal-structure fidelity of synthetic data using BGMM and standard Gaussian priors in Scenario~2.}
\label{tab:m2_causal}
\adjustbox{max width=\textwidth, center}{
\begin{threeparttable}
\begin{tabular}{lllrr}
\toprule
            Category &                        Metric &                                Direction &          BGMM &      Gaussian \\
\midrule
    Treatment Effect &                  CATE/ITE MAE &                                 ↓ better & 0.003 (0.000) & 0.004 (0.000) \\
    Treatment Effect &              CATE Correlation &                                 ↑ better & 0.971 (0.002) & 0.915 (0.004) \\
    Treatment Effect &                     ATE Error &                                 ↓ better & 0.000 (0.000) & 0.001 (0.000) \\
    Treatment Effect & TE Distribution Distance (W1) &                                 ↓ better & 0.001 (0.000) & 0.001 (0.000) \\
         Confounding &               Confounding MAE &                                 ↓ better & 0.002 (0.000) & 0.002 (0.000) \\
         Confounding &          Group-wise MAE (T=0) &                                 ↓ better & 0.002 (0.000) & 0.002 (0.000) \\
         Confounding &          Group-wise MAE (T=1) &                                 ↓ better & 0.002 (0.000) & 0.002 (0.000) \\
         Confounding &        Confounding Dist. (W1) &                                 ↓ better & 0.002 (0.000) & 0.002 (0.000) \\
   Overlap (decoder) &                           MSE &                                 ↓ better & 0.004 (0.000) & 0.005 (0.000) \\
   Overlap (decoder) &     Fraction within tolerance &                                 ↑ better & 0.997 (0.001) & 0.993 (0.001) \\
Overlap (propensity) &           Propensity AUC$^{a}$&                                       NA & 0.561 (0.007) & 0.563 (0.008) \\
Overlap (propensity) & Histogram overlap coefficient &                                 ↑ better & 0.850 (0.021) & 0.855 (0.015) \\
Overlap (propensity) &       Common support fraction &                                 ↑ better & 1.000 (0.000) & 1.000 (0.000) \\
Overlap (propensity) & Common support fraction (T=0) &                                 ↑ better & 1.000 (0.000) & 1.000 (0.000) \\
Overlap (propensity) & Common support fraction (T=1) &                                 ↑ better & 1.000 (0.000) & 1.000 (0.000) \\
\bottomrule
\end{tabular}

\begin{tablenotes}[flushleft]
\footnotesize
\item[a] AUC closer to 0.5 indicates stronger overlap.
\end{tablenotes}
\end{threeparttable}
}
\end{table*}

\begin{table*}[h]
\centering
\caption{Causal-structure fidelity of synthetic data using BGMM and standard Gaussian priors in Scenario~3.}
\label{tab:m3_causal}
\adjustbox{max width=\textwidth, center}{
\begin{threeparttable}
\begin{tabular}{lllrr}
\toprule
            Category &                        Metric &                                Direction &          BGMM &      Gaussian \\
\midrule
    Treatment Effect &                  CATE/ITE MAE &                                 ↓ better & 0.004 (0.000) & 0.004 (0.000) \\
    Treatment Effect &              CATE Correlation &                                 ↑ better & 0.956 (0.003) & 0.931 (0.005) \\
    Treatment Effect &                     ATE Error &                                 ↓ better & 0.001 (0.000) & 0.002 (0.000) \\
    Treatment Effect & TE Distribution Distance (W1) &                                 ↓ better & 0.002 (0.000) & 0.002 (0.000) \\
         Confounding &               Confounding MAE &                                 ↓ better & 0.004 (0.000) & 0.005 (0.000) \\
         Confounding &          Group-wise MAE (T=0) &                                 ↓ better & 0.005 (0.000) & 0.005 (0.000) \\
         Confounding &          Group-wise MAE (T=1) &                                 ↓ better & 0.003 (0.000) & 0.004 (0.000) \\
         Confounding &        Confounding Dist. (W1) &                                 ↓ better & 0.002 (0.000) & 0.002 (0.000) \\
   Overlap (decoder) &                           MSE &                                 ↓ better & 0.004 (0.000) & 0.300 (0.091) \\
   Overlap (decoder) &     Fraction within tolerance &                                 ↑ better & 0.996 (0.001) & 0.934 (0.004) \\
Overlap (propensity) &          Propensity AUC$^{a}$ &                                       NA & 0.889 (0.006) & 0.859 (0.005) \\
Overlap (propensity) & Histogram overlap coefficient &                                 ↑ better & 0.243 (0.012) & 0.309 (0.008) \\
Overlap (propensity) &       Common support fraction &                                 ↑ better & 0.998 (0.003) & 1.000 (0.000) \\
Overlap (propensity) & Common support fraction (T=0) &                                 ↑ better & 1.000 (0.000) & 1.000 (0.000) \\
Overlap (propensity) & Common support fraction (T=1) &                                 ↑ better & 0.996 (0.004) & 1.000 (0.000) \\
\bottomrule
\end{tabular}

\begin{tablenotes}[flushleft]
\footnotesize
\item[a] AUC closer to 0.5 indicates stronger overlap.
\end{tablenotes}
\end{threeparttable}
}
\end{table*}

\begin{figure}[t]
\centering
\includegraphics[width=\textwidth]{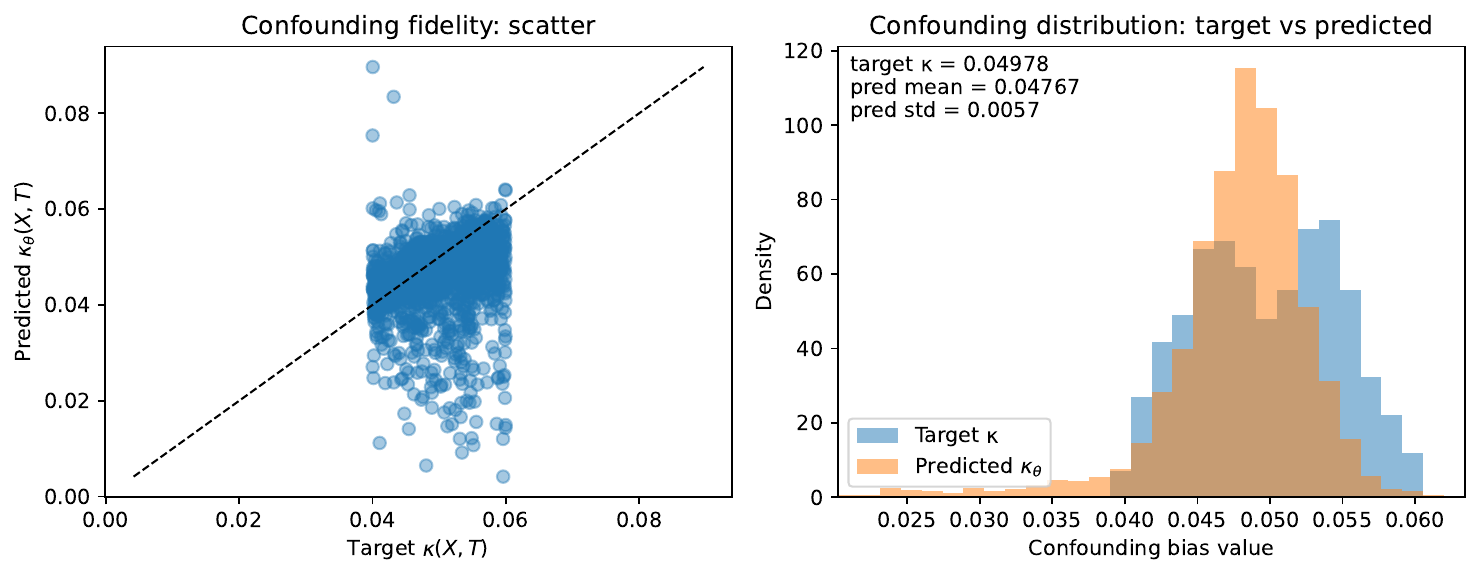}
\caption{Comparison of generated and target unmeasured confounding of the mcrpc dataset from CausalMix with a BGMM prior in Scenario~3.}
\label{fig:m3_bias}
\end{figure}


\begin{figure}[t]
\centering

\begin{subfigure}{0.48\textwidth}
  \centering
  \includegraphics[width=\linewidth]{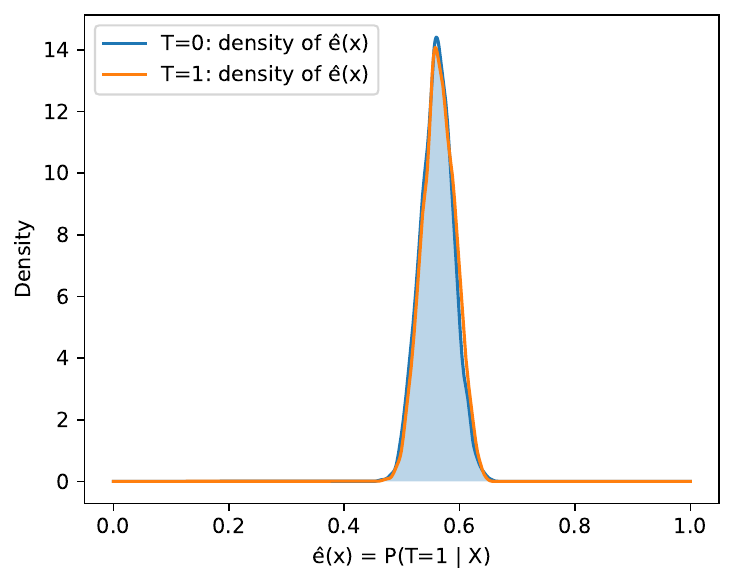}
  \caption{Scenario~1}
\end{subfigure}
\hfill
\begin{subfigure}{0.48\textwidth}
  \centering
  \includegraphics[width=\linewidth]{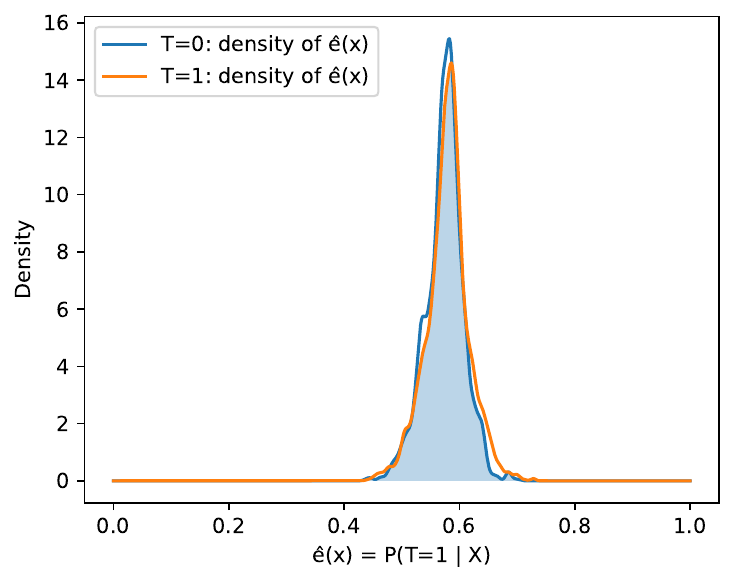}
  \caption{Scenario~2}
\end{subfigure}

\vspace{0.5cm}

\begin{subfigure}{0.6\textwidth}
  \centering
  \includegraphics[width=\linewidth]{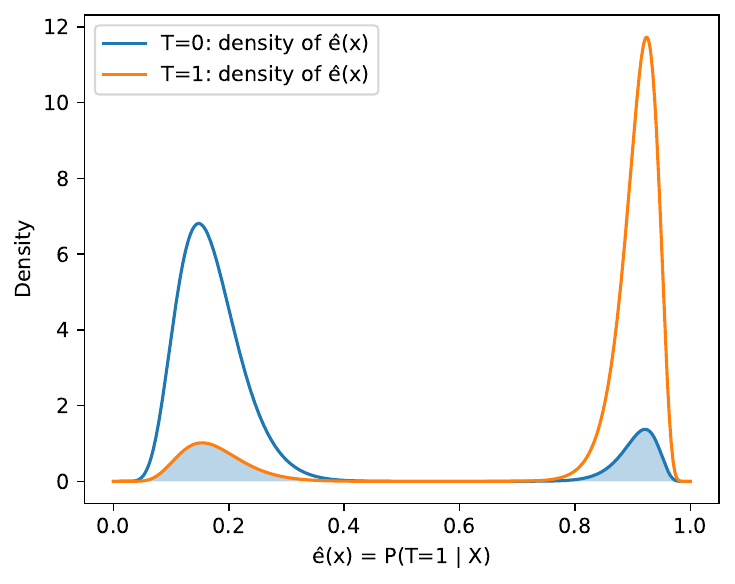}
  \caption{Scenario~3}
\end{subfigure}

\caption{Generated overlap of the mcrpc datasets from CausalMix with a BGMM prior across causal scenarios. Scenario 1: perfect overlap $\log\alpha(X)=0$; Scenario 2: moderate constant overlap $\log\alpha(X)=1$; Scenario 3: covariate-dependent overlap $\log\alpha(X)=2\big(2\,\text{Abiraterone\_prev}-1\big)$.}
\label{fig:all_overlap}
\end{figure}

\begin{table}[t]
\centering
\caption{Privacy of synthetic data using BGMM and standard Gaussian priors in Scenario~1.}
\label{tab:m1_privacy}
\adjustbox{max width=\columnwidth}{
\begin{tabular}{llrr}
\toprule
                     Metric & Direction &          BGMM &      Gaussian \\
\midrule
        Protection Fraction &  ↑ better & 0.595 (0.009) & 0.696 (0.008) \\
      Distance Ratio (mean) &  ↑ better & 2.535 (0.083) & 3.254 (0.163) \\
        Distance Ratio (p5) &  ↑ better & 0.233 (0.028) & 0.501 (0.003) \\
       Distance Ratio (p50) &  ↑ better & 1.039 (0.006) & 1.122 (0.006) \\
       Distance Ratio (p95) &  ↑ better & 4.841 (0.263) & 6.150 (0.304) \\
Standardized Distance Ratio &  ↑ better & 0.276 (0.003) & 0.333 (0.003) \\
\bottomrule
\end{tabular}

}
\end{table}

\begin{table}[t]
\centering
\caption{Privacy of synthetic data using BGMM and standard Gaussian priors in Scenario~2.}
\label{tab:m2_privacy}
\adjustbox{max width=\columnwidth}{
\begin{tabular}{llrr}
\toprule
                     Metric & Direction &          BGMM &      Gaussian \\
\midrule
        Protection Fraction &  ↑ better & 0.607 (0.007) & 0.747 (0.007) \\
      Distance Ratio (mean) &  ↑ better & 2.787 (0.193) & 3.723 (0.249) \\
        Distance Ratio (p5) &  ↑ better & 0.255 (0.021) & 0.624 (0.036) \\
       Distance Ratio (p50) &  ↑ better & 1.057 (0.007) & 1.195 (0.009) \\
       Distance Ratio (p95) &  ↑ better & 4.623 (0.207) & 7.681 (0.622) \\
Standardized Distance Ratio &  ↑ better & 0.282 (0.002) & 0.373 (0.007) \\
\bottomrule
\end{tabular}

}
\end{table}

\begin{table}[t]
\centering
\caption{Privacy of synthetic data using BGMM and standard Gaussian priors in Scenario~3.}
\label{tab:m3_privacy}
\adjustbox{max width=\columnwidth}{
\begin{tabular}{llrr}
\toprule
                     Metric & Direction &          BGMM &       Gaussian \\
\midrule
        Protection Fraction &  ↑ better & 0.720 (0.006) &  0.839 (0.004) \\
      Distance Ratio (mean) &  ↑ better & 3.298 (0.245) &  4.287 (0.089) \\
        Distance Ratio (p5) &  ↑ better & 0.396 (0.030) &  0.788 (0.014) \\
       Distance Ratio (p50) &  ↑ better & 1.169 (0.007) &  1.337 (0.009) \\
       Distance Ratio (p95) &  ↑ better & 6.046 (0.055) & 11.740 (0.493) \\
Standardized Distance Ratio &  ↑ better & 0.297 (0.003) &  0.559 (0.005) \\
\bottomrule
\end{tabular}

}
\end{table}

\begin{table*}[h]
\centering
\caption{Accuracy and coverage of causal estimators under no unmeasured confounding}
\label{tab:est_coverage}
\adjustbox{max width=\textwidth, center}{
\begin{tabular}{lrrrrr}
\toprule
                 Estimator & Runtime (s , SD) &  ATE RMSE &  Mean ATE SE &  ATE Coverage &  Mean CATE Coverage \\
\midrule
    Bayesian Causal Forest &      258.7 (9.5) &     0.010 &        0.010 &          0.90 &                0.91 \\
             Causal Forest &       25.1 (1.7) &     0.009 &        0.009 &          0.94 &                0.87 \\
       Standard Linear DML &        2.2 (0.1) &     0.009 &        0.009 &          0.96 &                0.90 \\
                 Lasso DML &     229.9 (10.1) &     0.009 &        0.008 &          0.88 &                0.37 \\
                   GBR DML &     268.1 (10.7) &     0.010 &        0.113 &          1.00 &                0.82 \\
Standard Linear DR-Learner &        2.3 (0.1) &     0.010 &        0.046 &          1.00 &                0.92 \\
          Lasso DR-Learner &     234.5 (10.6) &     0.010 &        0.009 &          0.88 &                0.40 \\
            GBR DR-Learner &     274.0 (10.8) &     0.010 &        0.125 &          1.00 &                0.82 \\
           Lasso X-Learner &      118.4 (4.1) &     0.024 &        0.036 &          1.00 &                0.82 \\
             GBR X-Learner &      160.4 (4.4) &     0.024 &        0.085 &          1.00 &                0.82 \\
\bottomrule
\end{tabular}

}
\end{table*}

\begin{table*}[h]
\centering
\caption{Accuracy and coverage of causal estimators under small unmeasured confounding}
\label{tab:est_coverage_biased}
\adjustbox{max width=\textwidth, center}{
\begin{tabular}{lrrrrr}
\toprule
                 Estimator & Runtime (s , SD) &  ATE RMSE &  Mean ATE SE &  ATE Coverage &  Mean CATE Coverage \\
\midrule
    Bayesian Causal Forest &      260.6 (9.6) &     0.023 &        0.011 &          0.48 &                0.86 \\
             Causal Forest &       25.3 (1.8) &     0.027 &        0.009 &          0.18 &                0.80 \\
       Standard Linear DML &        2.1 (0.1) &     0.026 &        0.009 &          0.24 &                0.84 \\
                 Lasso DML &      225.0 (7.9) &     0.025 &        0.010 &          0.28 &                0.26 \\
                   GBR DML &      262.8 (7.7) &     0.026 &        0.125 &          1.00 &                0.83 \\
Standard Linear DR-Learner &        2.2 (0.1) &     0.028 &        0.046 &          1.00 &                0.85 \\
          Lasso DR-Learner &      230.5 (8.4) &     0.028 &        0.008 &          0.14 &                0.25 \\
            GBR DR-Learner &      269.7 (7.3) &     0.028 &        0.132 &          1.00 &                0.81 \\
           Lasso X-Learner &      116.4 (3.8) &     0.008 &        0.035 &          1.00 &                0.86 \\
             GBR X-Learner &      158.6 (4.3) &     0.009 &        0.088 &          1.00 &                0.83 \\
\bottomrule
\end{tabular}

}
\end{table*}
\begin{figure*}[t]
\centering
\includegraphics[width=\textwidth]{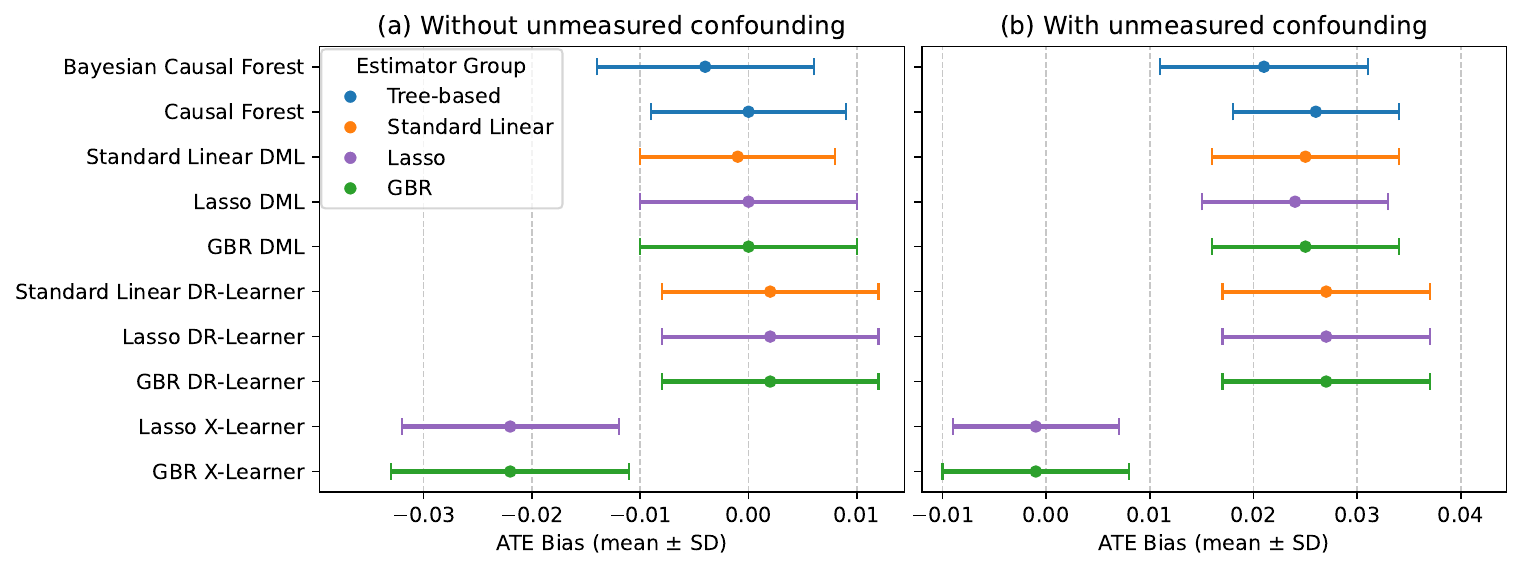}
\caption{Comparison of ATE bias across estimators under no unmeasured confounding ($\kappa=0$) and moderate unmeasured confounding ($\kappa=0.02$). Points denote mean bias across 50 independent replications and error bars indicate one standard deviation. Estimator groups are defined based on the CATE model.}
\label{fig:app1:ate_bias}
\end{figure*}

\begin{figure*}[t]
\centering
\includegraphics[width=0.8\textwidth]{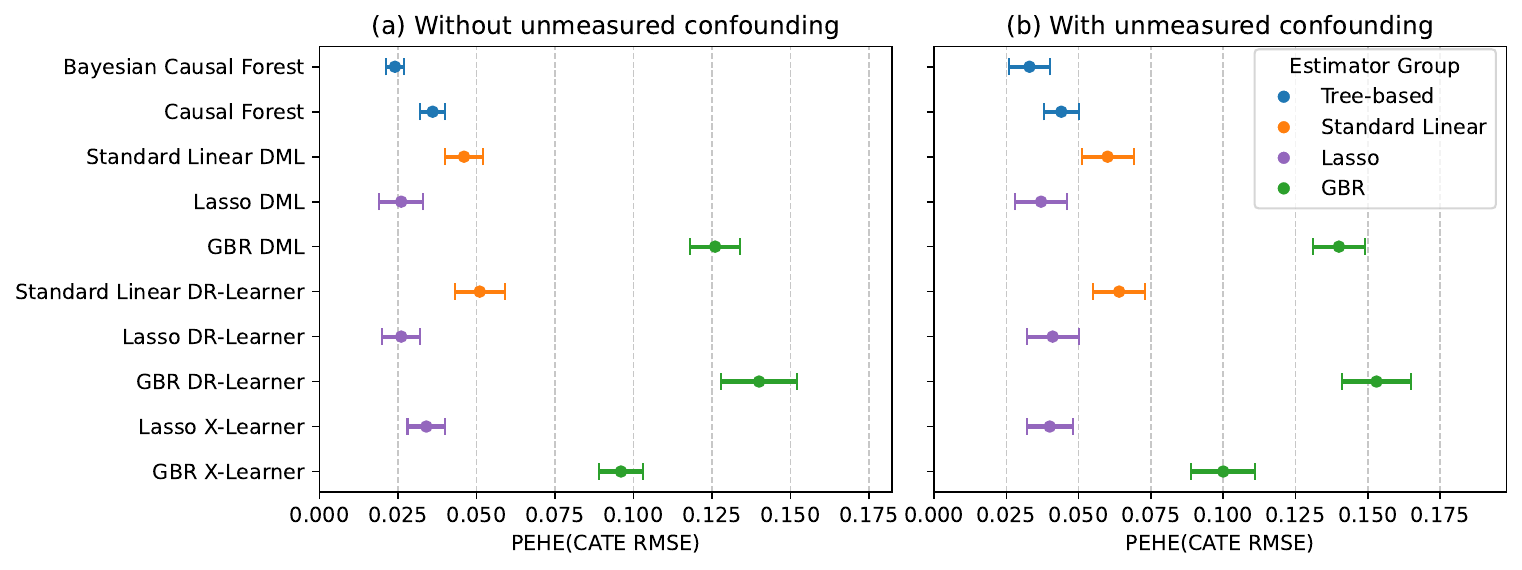}
\caption{Comparison of PEHE across estimators under no unmeasured confounding ($\kappa=0$) and moderate unmeasured confounding ($\kappa=0.02$). Points denote mean RMSE across 50 independent replications and error bars indicate one standard deviation. Estimator groups are defined based on the CATE model.}
\label{fig:app1:cate_rmse}
\end{figure*}

\begin{figure*}[t]
\centering
\includegraphics[width=0.8\textwidth]{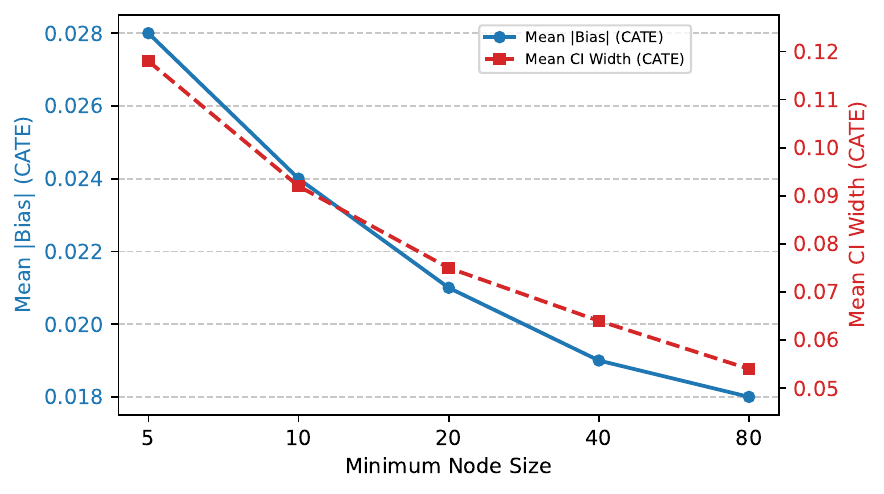}
\caption{Sensitivity of CATE accuracy and uncertainty to the minimum leaf size in causal forests (2,000 trees). The solid curve shows mean absolute CATE bias and the dashed curve shows mean CATE confidence interval width, averaged over 50 replications.}
\label{fig:app2:bias_ciw}
\end{figure*}

\end{document}